\documentclass[aps,prx,superscriptaddress,twocolumn]{revtex4}
\usepackage{color}
\usepackage{graphicx}
\usepackage{times}
\usepackage[mathscr]{euscript}
\usepackage{verbatim}
\usepackage{amsmath}
\usepackage{amsthm}
\usepackage{amssymb}
\usepackage{amsbsy}
\usepackage{tabularx}
\usepackage{natbib}
\newcommand{\non}{\nonumber}

\newcommand*{\MyPath}{/home/amit/Documents/Projects/DisCondDOS/Numerics/results}
\newcommand*{\MyNewPath}{/home/amit/Documents/Projects/DisCondDOS/Numerics}
\DeclareMathOperator{\sgn}{sgn}

\newcommand{\mathbbm}[1]{\text{\usefont{U}{bbm}{m}{n}#1}} 

\begin{document}

\title{How to infer non-Abelian statistics and topological visibility from 
tunneling conductance properties of realistic Majorana nanowires}
\author{S. Das Sarma}
\author{Amit Nag}
\author{Jay D. Sau}
\affiliation{Condensed Matter Theory Center and Joint Quantum Institute and Station Q Maryland, Department of Physics, 
University of Maryland, College Park, Maryland 20742-4111, USA}

\date{\today}

\begin{abstract}
We consider a simple conceptual question with respect to Majorana zero modes in semiconductor 
nanowires:  Can the measured non-ideal values of the zero-bias-conductance-peak in the tunneling 
experiments be used as a characteristic to predict the underlying topological nature of 
the proximity induced nanowire superconductivity?  In particular, 
we define and calculate the topological visibility, 
which is a variation of the topological invariant associated 
with the scattering matrix of the system as well as the 
zero-bias-conductance-peak heights in the tunneling measurements, in the presence of dissipative 
broadening, using precisely 
the same realistic nanowire parameters to connect the topological invariants with the 
zero bias tunneling conductance values.  This dissipative broadening is present in both (the existing) tunneling 
measurements and also (any future) braiding experiments as an inevitable consequence of a finite braiding 
time.
The connection between the topological visibility and the conductance allows us to obtain the visibility 
of realistic braiding experiments in nanowires, and to conclude that the current experimentally 
accessible systems with non-ideal zero bias conductance peaks may indeed manifest (with rather low visibility) 
non-Abelian statistics for the Majorana zero modes.  
In general, we find that large (small)  
superconducting gap (Majorana peak splitting)
is essential for the manifestation of the non-Abelian braiding 
statistics, and in particular, a zero bias conductance value 
of around half the ideal quantized Majorana value should be 
sufficient for the manifestation of non-Abelian statistics 
in experimental nanowires.
Our work also establishes that as a matter of principle the topological
transition associated with the emergence of Majorana zero modes in finite nanowires 
is always a crossover (akin to a quantum phase transition at finite temperature) 
requiring the presence of dissipative broadening (which must be larger than the Majorana 
energy splitting in the system) in the system.  For braiding, this dissipation is supplied by 
the finite speed of the braiding process itself  
which must be diabatic in any real experiment for braiding to succeed.

\end{abstract}

\maketitle
\section{Introduction}
In 2012, in an important experimental report~\cite{Mou} Mourik \textit{et. al.}
presented evidence for the possible existence of non-Abelian Majorana 
zero modes (MZMs) in InSb nanowires subjected to an external magnetic (Zeeman)
field in proximity to an ordinary \textit{s}-wave superconductor (NbTiN in Ref.~\cite{Mou}).
This experiment followed precisely earlier theoretical predictions 
~\cite{Jay,Rom,Jas1,Jay1,Yuv}
on how to create, localize, and observe MZMs in nanowires 
by judiciously combining 
Rashba spin-orbit coupling, Zeeman spin splitting, and \textit{s}-wave superconducting 
proximity effect--- the basic idea of the prediction, first explicitly developed 
in Ref.~\cite{Jay}, being that a combination of spin-orbit coupling and spin splitting could, 
in principle, convert ordinary \textit{s}-wave superconductors into topological 
(effectively spinless, although spin could play an important 
role in some situations~\cite{liu2015universal}) 
\textit{p}-wave superconductors which would carry localized MZMs at 
suitable defect sites (including the boundaries) provided the Zeeman field is large 
enough to overcome the \textit{s}-wave superconducting gap,
thus inducing a chiral~\cite{Jay,Jay1} (or helical~\cite{Rom,Yuv})
topological \textit{p}-wave gap.
The observation by Mourik \textit{et. al.} received considerable support from several 
other independent experiments~\cite{Den,Leo,Das,Fin,Chu,Cha}
in semiconductor nanowires (both InSb and InAs)
where signatures for the existence of MZMs were reported by other groups.  
The fact that a spinless \textit{p}-wave superconducting wire would have 
localized MZMs (with non-Abelian anyonic braiding statistics) at the wire ends was already 
pointed out by Kitaev almost 15 years ago~\cite{Kit}, and Sengupta \textit{et. al.}~\cite{Sen},
also 15 years ago by coincidence, established the 
possibility that these MZMs, localized as zero-energy midgap 
interface states 
in topological superconductors,
could be experimentally detected using differential tunneling 
spectroscopy where the perfect Andreev reflection associated with the MZMs would produce a quantized
zero bias conductance (precisely at zero energy, i.e. at midgap) which would be a signature 
of their existence.  A similar signature for Majorana fermion edge
states at the interface of a superconductor and surface of a topological insulator was made by Law 
\textit{et. al.}~\cite{law2009majorana}.
Such  zero bias conductance peaks (ZBCP), also predicted in the specific
context of the semiconductor nanowires by Sau \textit{et. al.}~\cite{Jay1}, are precisely the 
observations of most of the experimental claims~\cite{Mou,Den,Leo,Das,Fin,Chu,Cha}
for the possible observation of MZMs in
semiconductor nanowires.  The subject has created enormous excitement in the community because 
of the novelty associated with topological superconductivity and non-Abelian statistics as well as
the possibility of carrying out fault-tolerant topological quantum computation using MZMs~\cite{Che}, 
and has been extensively reviewed in the recent literature 
~\cite{Jas,Bee1,Kar,Mar,Sta,San,sarma2015majorana}.
The current work provides a link between the experimentally observed ZBCP in 
the semiconductor nanowires and the possible topological 
properties of the underlying MZMs through extensive numerical simulations 
calculating certain topological invariants along with ZBCP values
for the same nanowire parameters.
 
Although the MZM interpretation (i.e. the relevant semiconductor nanowires in these 
experiments carry non-Abelian MZMs at the ends of the wires) 
is the most natural explanation for the experimental observations
~\cite{Mou,Den,Leo,Das,Fin,Chu,Cha}  in the semiconductor nanowires, particularly 
in view of the existing theoretical predictions
~\cite{Jay,Rom,Jas1,Jay1,Yuv} preceding the experiments, many questions remain, 
and the possibility that the ZBCPs arise
from some other non-topological mechanism cannot definitively be ruled out.  
In this regard several mechanisms have been suggested which do not necessitate the 
occurrence of a topological phase transition (TPT) for the appearance of ZBCP~\cite{Alt1,
Pat,Edu,Fal,Kel,Jie,pikulin2012weakanti}. None of these alternate non-topological 
mechanisms requires the ZBCP to be canonically quantized to $2e^2/h$ 
as is the expected ideal value for the perfect Andreev reflection associated with 
MZMs~\cite{Sen,flensberg2010tunneling,law2009majorana,wimmer2011quantum}.
But the observed ZBCP in the experiments is not quantized either 
and is, in fact, typically well below $2e^2/h$, 
which has been explained as arising from a 
number of physical mechanisms affecting the experimental situation~\cite{lin2012zero}. Several
experimentally observed features of the ZBCP fall short of the idealized theoretical
expectations: 1) ZBCP height
is much smaller than the predicted perfect quantized
value of $2e^2/h$; 2) Recent attempts to produce ZBCPs 
closer to the quantized value typically appear to lead to  a  
broadening much larger than the thermal value \cite{delftballistic};
3) ZBCP often does
not manifest the expected oscillatory peak splitting 
(with increasing Zeeman field) predicted for MZMs
~\cite{Meng,Meng1,Rai,Das1} in finite length wires where the MZMs at the two wire ends should 
overlap with each other
producing an energy splitting around zero bias. 
In spite of a large number of theoretical investigations in the literature
explaining the observed non-ideal ZBCP properties as a result of 
underlying Majorana modes and 
its interplay with
disorder, temperature, lead couplings and other non-idealities
~\cite{lin2012zero,takei2013soft,SauSarma,
brouwer2011topological,Pie,prada2012transport,law2009majorana,chevallier2013andreev},
it is hard to discount alternate non-topological 
mechanisms conclusively yet, particularly since the experimentally observed ZBCP 
remains well below the ideal quantized value of $2e^2/h$.
At best, the experiments seem to (weakly) satisfy
the necessary conditions for MZMs (i.e. the observation of a ZBCP under the predicted conditions),
but not the sufficient conditions for claiming conclusive evidence supporting the existence of 
non-Abelian MZMs.  It is entirely possible, perhaps even likely, that the invariable existence of
finite disorder, finite temperature, finite wire length, finite coupling to the tunneling leads, 
imperfect proximity coupling, and other non-idealities in the realistic systems make it completely
impossible to observe the predicted perfect ZBCP quantization of $2e^2/h$ in the experimental setups.
It is encouraging that recent materials improvements in making the nanowires have led to a substantial 
enhancement in the observed value of the ZBCP although 
it is still smaller than the
perfect quantized value of $2e^2/h$~\cite{MarUnp}.
However, it should be noted that zero bias tunneling conductance peaks 
could arise in superconductors from a multitude of reasons, 
and cannot by itself be taken as compelling evidence for the existence of MZMs.
We need some direct evidence for the topological phase 
transition accompanying 
the emergence of 
MZMs~\cite{lobos2015tunneling,fregoso2013electrical}    
and some measurements for topological properties.
Perhaps the controversy regarding the existence or not of  
MZMs in nanowires would not arise if every experimental detection 
of the ZBCP found a value close to the expected universal quantized
Majorana value of $2e^2/h$, but the fact that the experimental ZBCP value 
is both nonuniversal from experiment to experiment and is always much 
lower than $2e^2/h$ casts a dark shadow on the MZM interpretation 
of the experimental tunneling transport measurements.

Given that the defining exclusive property of the MZMs is 
their topological non-Abelian braiding 
characteristics~\cite{Rea,Iva} with the MZMs being subgap 
zero-energy non-Abelian anyonic excitations, 
it would seem that an experiment conclusively establishing
their non-Abelian character would be the decisive sufficient condition for their existence.  
Indeed several proposals have been put forth in the literature for probing the 
non-Abelian braiding properties of 
MZMs~\cite{Jas2,sau2011QComp,clarke2011QComp,hyart2013flux,sarma2015majorana,
amorim2015majorana,van2012coulomb,liang2012manipulation,romito2012manipulating,
halperin2012adiabatic,kotetes2013engineering,clarke2015bell,liu2013manipulating,
chiu2015majorana,sau2011controlling,sau2010universal}, 
and experimental efforts are 
currently underway to carry out MZM braiding to test their 
non-Abelian properties.  An observation of the non-Abelian braiding properties would go a long way in
establishing the existence of true MZMs in nanowires.  The current work is a theoretical attempt to
directly test what such a non-Abelian braiding experiment is likely to observe in realistic nanowires 
where the ZBCP is very far from being quantized and has large broadening.
We establish in this work a clear connection between the observation 
of imperfect ZBCP and underlying topological properties, showing that 
the current experimental observations are indeed (but only marginally so) 
consistent with the 
possibility of the nanowires hosting non-Abelian Majorana zero modes purely 
from the perspective of braiding-related topological properties.
 
To provide a context, we start
by assuming that the experimental observation in Ref.~\cite{Mou} (and other nanowire experiments) of the ZBCP 
is indeed a signature of (highly) imperfect MZMs which, because of various non-idealities
in the system (e.g. disorder, temperature, tunnel 
coupling to the environment, finite wire length, 
Majorana splitting, etc.), produce a ZBCP which is highly suppressed (and broadened) compared 
to the
canonically quantized value of $2e^2/h$~\cite{lin2012zero,takei2013soft,
brouwer2011topological,prada2012transport,law2009majorana}.
The immediate question then is whether (or perhaps, 
to what extent) such imperfect almost-MZMs would  have intrinsic non-Abelian
braiding properties possibly showing up experimentally (or numerically in our study). 
In the absence of a braiding experiment to directly observe non-Abelian statistics for Majorana 
exchange at present, we are left to speculate on the extent to which non-Abelian
statistics would be observed when nanowire 
MZMs are braided based on the only available experimental
signal for their existence, i.e., ZBCP. It is then prudent to ask if we can relate the observed (non-ideal) 
characteristics
of the ZBCP, i.e. height and width of the peak, to the topological 
content of the approximate MZMs. 
Our work quantitatively establishes this connection and hence sheds light on the possibility of 
observing the topological
nature of MZMs (in terms of non-Abelian exchange statistics) in future braiding experiments
carried out in the same (or similar) samples as the ones 
currently manifesting non-ideal ZBCPs.
Thus, rather than simulating a future braiding experiment,  
we look at the electron tunneling properties  
of the nanowire close to zero energy 
to answer the extent
to which it might be possible to demonstrate the non-Abelian characteristics of the Majorana modes
for the given set of physical quantities of the system viz. Majorana splitting, topological gap, tunneling strength  etc.

Kitaev suggested
calculating a precisely defined quantity--- 
topological invariant(TI)--- to distinguish between 
trivial and topological
phases in a \textit{p}-wave superconducting wire~\cite{Kit}. The invariant
suggested by Kitaev is suitable for systems with periodic boundary conditions.
An appropriate generalization 
of the TI suitable for a finite system with an open
boundary condition was introduced by 
Akhmerov \textit{et. al.}~\cite{Akh} in terms of the S-matrix of the associated
system. Since we want to relate features observed in tunneling experiments
(which are necessarily conducted on finite wires with open boundary 
conditions) to the underlying
topological nature of the MZMs, we would use the proposed 
scattering matrix invariant to calculate the TI of the realistic system in order 
to quantify the topological
nature of the semiconductor nanowire as we 
tune the physical parameters that are critical to the existence
of MZMs, viz., wire length and Zeeman field.
In fact a variant of the scattering matrix TI
were recently used in part
by Adagedeli \textit{et. al.}~\cite{adagideli2014effects} to show the 
existence of topological phase for  
disordered (mean free path $L_{mf}$ shorter than the induced coherence length $\xi$) 
semiconductor nanowires.

However, a subtlety of the usual definition of the 
TI $Q_0=\mathrm{sgn}[\mathrm{det}(r)]$, where $r$ is the reflection matrix from the 
end is that it requires us to ignore transmission of quasiparticles in-between 
the ends of the wire~\cite{Ful,Akh,adagideli2014effects}. Such transmission of 
quasiparticles always exist for the finite wires we consider in this work. 
In fact, as we will discuss in more detail in this work, for finite wires 
the TI $Q_0$ is always trivial when one uses the exact reflection 
matrix (as opposed to the effectively semi-infinite approximation 
used in Refs.~\cite{Akh,adagideli2014effects}).
In this work, instead of ignoring transmission across the wire, 
we circumvent this problem by introducing dissipation into the system.
While some form of dissipation has been important
in previous calculations of the scattering
matrix TI~\cite{Ful,adagideli2014effects}, dissipation in our work represents the finite rate 
of braiding.  
As pointed out in previous works~\cite{pikulin2012weakanti,pikulin2012topological}
dissipation can change the qualitative behavior of Majorana modes 
and the TI. 

The standard scattering TI 
$Q_0$ is not sensitive to imperfections of the topological phase such as
transmission of quasiparticles through the wire. Such transmission through
the wire would interfere with topological signatures of Majorana modes such 
as conductance quantization and non-Abelian statistics. To remedy this, 
we define a variant of the TI, $Q=\mathrm{det}(r)$, which we refer 
to as topological visibility (TV), as a measure of the topological character 
of the system. From the calculations presented here, it will become clear that
the TV is better suited to determining the visibility of signatures of the 
Majorana fermion such as quantized conductance peak and non-Abelian statistics
than the TI, which is just the sign of the TV.  
In the limit that we ignore transmission through the wire so that $r$ is unitary,
which is the case considered in Refs.~\cite{Akh,adagideli2014effects,Ful}, 
this quantity is 
identical to $Q_0$. 
One might be concerned that the topological
visibility, $Q$, is not quantized as $Q_0$. However, $Q$ is quantized as long as 
the system is properly gapped so that $r$ is unitary. Whenever $Q$ is not quantized, 
which is near a topological phase transition, whether $Q_0$ is trivial or not depends
on non-universal details of the system which determine whether $\mathrm{det}(r)$ 
is slightly
positive or slightly negative. To keep our terminology consistent with previous
works~\cite{Akh,adagideli2014effects,Ful}, we will refer to $Q<0$ to 
be topological (i.e. $Q_0=\mathrm{sgn}(Q)=-1$)  and $Q>0$ to be non-topological. 
The presence of dissipation eliminates
the discreteness of the topological visibility $Q$ 
by relaxing the unitarity of the theory, leading to the possibility 
of the TV being any number between +1 and -1 
instead of having a magnitude precisely equal to unity.
Only when $Q$ is close to it's extreme values $\pm 1$ can $Q_0$
be reliably determined to be topological or not. 
The competition 
between the strength of dissipation and the finite size
splitting of Majorana modes in determining the TV is the
central focus of our work. In fact, our work establishes that the emergence
of MZMs in any finite length wires (i.e. in any experimental system which 
must always use finite wires) is always a 'topological quantum crossover' 
rather than a 'topological quantum phase transition' where dissipative 
broadening plays a fundamental role--- rather trivially, there is no 
topological phase in the absence of broadening in any finite length wire 
since the MZMs are never precisely at zero energy in finite wires!  The TV 
of the finite system taking on any possible value between +/- 1 rather than
being precisely equal to +1 (non-topological) or -1 (topological) is a direct 
consequence of the topological transition being a crossover in the finite system 
with broadening-- without any broadening, the finite system must always by 
definition have a TV equal to 1 since the MZM is always displaced from the 
energy zero.  We identify the topological crossover point as the TV passing 
through zero in our calculation with the TV $< (>)$ 0 being identified as the 
topological (non-topological) phase.  We also identify the deviation in the 
magnitude of the TV from unity being the direct manifestation of finite 
'visibility' in the braiding experiment-- closer the TV is to unity in 
magnitude, higher is the visibility for the corresponding phase 
(depending on whether the TV is positive or negative).

The dissipation we introduce is not just a mathematical convenience
and is an actual physical quantity present in the real experimental nanowires.
Dissipation can play a role in reducing the conductance from the 
quantized Majorana value to the experimentally observed value (even at zero temperature). 
Similarly, dissipation
might be responsible for increasing the width of the zero-bias peak beyond 
the thermal width\cite{delftballistic}. One might wonder if it is possible for the 
dissipative broadening to exceed temperature. In fact, coupling to a fermion 
bath can lead to such dissipation even at nearly zero temperature. 
Such a fermion bath can arise from subgap states at the semiconductor-superconductor
interface generated by disorder in the superconductor \cite{cole2016proximity}.
While a finite array of such subgap states is usually coherent, the presence of 
weak interaction and temperature coupled to such subgap states can lead to 
decoherence of the fermions, turning such a large density of subgap states into 
a fermion bath. Alternatively such a fermion bath can arise from subgap states 
in vortices generated by the magnetic field.

We will relate the topological nature of MZMs calculated in the tunneling
conductance setup to the non-Abelian braiding statistics of MZMs
through the appropriate direct numerical calculations of both the ZBCP and 
TV magnitudes in realistic systems, establishing correlations among them.
If the experimentally observed ZBCPs are indeed almost-MZMs (and not spurious effects arising from
totally distinct mechanisms that have nothing to do with topological superconductivity), then our work would 
provide a useful guide for the expected visibility of a non-Abelian braiding experiment in real samples
since we start by numerically calculating ZBCPs in the nanowires 
ensuring that the calculated ZBCP magnitudes are approximately consistent with
experimental observations.  Our work in fact encompasses two qualitatively distinct realistic aspects
of the experimental situation.  First, we establish the quantitative connection between having a ZBCP 
strongly suppressed from the quantized $2e^2/h$ value and the topological content of the associated almost-MZMs,
i.e., we investigate how suppressed the ZBCP could be from $2e^2/h$ and still manifest some topological character.
Second, we investigate the deleterious effects of MZM splitting, which must invariably be present in all finite
nanowires because of the overlap of the MZMs from the two ends, on the braiding properties
(or more precisely, on the value of the TV which distinguishes topological 
and trivial phases). The key concepts of dissipative broadening 
and realistic finite lengths of the nanowires hosting MZMs play crucial 
conceptual as well as quantitative roles in our theory.

The reason for focusing on the tunneling
scenario is twofolds. First, Majorana tunneling experiments have already been successfully conducted 
whereas the braiding experiments with nanowire Majorana are proposed future works. This
allows us to work with known experimental parameters and check our tunneling 
conductance results against the existing data that
is either published or in principle should be under present experimental reach. 
Therefore, quantitative expectations about putative non-Abelian Majorana braiding experiment 
of the future can be drawn based upon the available data and present experiments
on the tunneling conductance by relating both sets of results on the same system. 
Second, it is conceptually easier to characterize and computationally easier to numerically simulate electron tunneling into Majorana
nanowires than a braiding operation.
We point out that ramifications of braiding operations on MZMs in an
experiment have been studied theoretically and many detailed effects
and subtleties have been pointed out in Refs.~\cite{pedrocchi2015majorana,
karzig2013boosting,cheng2011nonadiabatic,
scheurer2013nonadiabatic,karzig2015optimal,karzig2015shortcuts}. However, since we are 
focusing on the topological content of stationary Majorana modes (probed numerically by
simulating a tunneling conductance measurement), our result would represent the 
best possible outcome one may hope to get towards observing the non-Abelian 
braiding statistics of MZMs.
In particular, our work specifically connects the outcome of a 
braiding experiment (i.e. the direct measurement of the TV in a system)
in relation to the measurement of the tunneling conductance in the same sample,
answering the question whether a given value of a measured (in our case, numerically)
ZBCP value is consistent or not with a topological value for the 
(numerically calculated) TV.
In general, the non-Abelian character in a Majorana braiding experiment 
will be observed for fast enough braiding operation 
so that the energy uncertainty
associated with the braiding time is larger than the 
Majorana splitting, which 
will entail
approximate Majorana modes to appear to be roughly degenerate 
(as opposed to being well-split).
However the experiment must distinguish the Majorana 
modes 
from the
continuum set of (above-gap) bulk states. 
Therefore, the the speed of the braiding must be slow enough so that the associated
energy uncertainty is not of the order of the topological gap. 
Or in other words, the braiding operation should be slow 
with respect to the inverse topological gap, but fast compared with 
the Majorana splitting.  
We argue that this is in complete analogy to how dissipative broadening, 
which is likely present in a tunneling conductance set-up, must be larger than Majorana splitting
but smaller than the topological gap to realize a nearly quantized conductance peak and also 
a topologically non-trivial value for the TV.
Our detailed numerical simulations quantify these conceptual points.
In fact, our work clearly establishes that one can make quantitative
statements about 'how topological' a particular system could be 
(at least, the upper bound) based 
simply on a detailed knowledge of the ZBCP peak height and broadening. 

In this work we explore the connection between conductance  
and the TV and calculate their dependence 
on the Majorana splitting and the energy gap.
These effects are studied for a specific Majorana hosting semiconductor
Rashba nanowire (e.g. InSb or InAs nanowire with strong Rashba spin-orbit coupling)
model proposed by Lutchyn \textit{et. al.} and 
Oreg \textit{et. al.}~\cite{Rom,Yuv}. 
This 
particular model benefits from having been studied extensively theoretically
(esp. see Refs.~\cite{SauSarma,sau2012experimental}) as well as from being
used as the theoretical guide to realize MZMs experimentally~\cite{Mou}. 
The paper is organized as follows. 
In Sec. II we introduce the model 
Hamiltonian and write it in its discretized form to make it amenable
to numerical techniques.
In Sec. III we investigate the 
effect of relevant energy scales, namely Majorana splitting and broadening,
on the behavior of the TV and conductance near the topological
phase transition
as well as deep in the topological phase. 
Particular emphasis is placed on possible correlations between the two quantities
in this general set-up.
In Sec. IV we use
the relationship between braiding, tunneling conductance,
and TV to study how conductance 
measurements can be used to characterize the outcomes of braiding experiments.
Finally, we 
conclude in Sec. V.
Appendix A reviews the details  of calculating the conductance and TV from the scattering matrix 
for the nanowire model obtained using KWANT~\cite{groth2014kwant}.
In Appendix B we discuss some  
more technical subtleties that arise in the numerical calculations using the S-matrix leading to the TV.

\section{Model Hamiltonian}
\begin{figure}
 \begin{center} 
\includegraphics[width=\columnwidth]{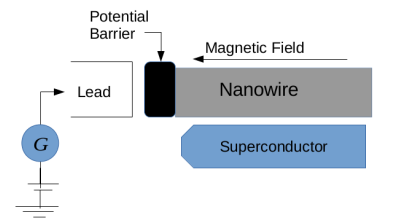}
\end{center}
\caption[Schematic for measuring tunneling conductance]{(Color Online)
A schematic diagram for measuring the tunneling conductance. 
One end of the Rashba nanowire is shown attached to a normal  lead. The 
lead is connected to the nanowire through a potential barrier. Magnetic field is parallel
to the nanowire. Proximate \textit{s}-wave superconductor is responsible 
for the superconducting 
order parameter in the nanowire.}
\label{cartoon}
\end{figure}
A schematic representation for an experimental setup to measure 
tunneling conductance is shown in Fig.~\ref{cartoon}.
A semiconductor nanowire  with Rashba spin-orbit coupling (SOC) 
is attached to a normal 
lead through a potential barrier localized at the end. A magnetic field is applied
parallel to the wire (perpendicular to the SOC direction) 
and an \textit{s}-wave superconductor is 
placed in proximity to the nanowire to facilitate
Cooper pair tunneling into the semiconductor effectively endowing the nanowire with 
an \textit{s}-wave superconducting order parameter through proximity coupling. 
A voltage bias $V$ is applied to the lead
and the tunneling 
current $I$ is measured to obtain the differential conductance 
\begin{align}
 G = dI/dV.
\end{align}
As discussed in more detail in the appendix, for $N$ conducting channels in the lead,
the conductance $G$ can be 
computed from the normal reflection matrix $r_{ee}$ and the Andreev reflection matrix 
$r_{eh}$ through the relation, 
\begin{align}
&G=N - \mathrm{Tr}(r_{ee}r_{ee}^{\dagger}-r_{eh}r_{eh}^{\dagger})
\end{align}
and the TV can be computed from the zero-frequency 
reflection matrices~\cite{Akh} as

\begin{align}
&Q=Det[r_{ee}r_{ee}^*-r_{ee}r_{eh}r_{ee}^{-1}r_{eh}^*].
\end{align}

The reflection matrices can be computed given the system and lead 
Hamiltonian, which we discuss in the remainder of the section.

Let us consider a particular semiconductor Rashba nanowire model introduced by
Lutchyn \textit{et. al.}~\cite{Rom}--- see also Refs. Sau \textit{et. al.}~\cite{Jay} 
and Oreg \textit{et. al.}~\cite{Yuv}--- which 
was shown to support 
MZMs at the two ends in the clean limit.
These theoretical works directly 
motivated the nanowire Majorana experiments
of Refs.~\cite{Mou,Den,Leo,Das,Fin,Chu,Cha}.
The 
BdG Hamiltonian describing the 1D nanowire in the presence of 
Rashba SOC, Zeeman splitting,
and superconducting proximity effect, is given by 
\begin{align}
  H_{sys} &= \left(-\frac{1}{2m^*}\partial_x^2 + i\alpha_R\sigma_y \partial_x - \mu
  \right)
 \tau_z \non \\ &+ \mu_0 B\sigma_x + \Delta_0\tau_x,
 \label{Hsys}
\end{align}
where, 
$m^*,\alpha_R,\mu$ and $\Delta_0$ are the effective mass, the strength of Rashba SOC, the
chemical potential and the proximity induced superconducting gap, respectively. 
Throughout this paper we set $\hbar = 1$.
Here and henceforth $\tau_{x,y,z}$ and $\sigma_{x,y,z}$  are Pauli matrices acting on particle-hole
and spin space, respectively.
$\mu_0 = g\mu_B$ is the usual gyromagnetic ratio times the Bohr magneton
defining the Zeeman field strength $\mu_0 B$.
To make it amenable to numerical
techniques we discretize the continuum Hamiltonian as,

\begin{align}
H_{sys}^{dis} &= \sum_{\pmb{n}=1}^{N}[-t \left(|n+1\rangle\langle n|
 + \mathrm{H.c.}\right)\tau_z \nonumber \\
&+ i\alpha \left(|n+1\rangle\langle n|
- \mathrm{H.c.}\right)\sigma_y\tau_z 
+\Delta_0|n\rangle\langle n| \tau_x \nonumber \\
&+ (-\mu +2t)|n\rangle\langle n|
 \tau_z + V_Z|n\rangle\langle n|\sigma_x],
\label{Hsys1}
\end{align}
where $t = \frac{1}{2m^*a^2}$ with $a$ being the lattice constant
for the discretized tight-binding model in Eq.~\eqref{Hsys1}. Length of the nanowire
is given by
$L = aN$ where $N$ is the number of unit cells in the wire, the SOC strength is given by
$\alpha = \frac{\alpha_R}{2a}$, and we have defined the Zeeman field strength, 
$V_Z \equiv \mu_0 B$ giving the spin splitting. 
The nondiagonal terms in the site index arise from nearest neighbor hopping.
This system has been shown to support 
MZMs~\cite{Rom,Yuv}. In fact, for a clean nanowire it is now well-known that MZMs 
exist as stable  localized zero energy excitations at 
the ends of the nanowire whenever $V_Z > \sqrt{\Delta_0^2+\mu^2}$. 

Before we describe the normal  leads that attach to the nanowire
to create the normal-superconductor (NS)
junction (see Fig.~\ref{cartoon}) for tunneling measurements, we first 
comment on an important quantity that can be calculated 
from the system Hamiltonian.
It is known that MZMs contribute a local density of states (LDOS) zero bias 
peak in the topological phase~\cite{SauSarma}. 
LDOS not only
probes the presence of zero energy modes, but also whether the zero energy
mode is localized close to the edge of the wire.
In fact,
computing or measuring the LDOS is the simplest probe to test the presence or absence
of MZMs in the system.
LDOS at a given energy
$\epsilon$ and site $i$ is given by
\begin{align}
 &\mathrm{LDOS}(\epsilon,i) =\\ \non & 
 \sum_n\left(|u_{n\uparrow}(i)|^2+|u_{n\downarrow}(i)|^2 
 +|v_{n\uparrow}(i)|^2+|v_{n\downarrow}(i)|^2\right)\delta(\epsilon-\epsilon_n),
\end{align}
where $\psi_n(i) = \left(u_{n\uparrow}(i),u_{n\downarrow}(i),v_{n\uparrow}(i),v_{n\downarrow}(i)\right)^T$
is the $i$-th component of eigenvector $\psi_n$ of the Hamiltonian matrix
$H_{sys}^{dis}$
with eigenvalue $\epsilon_n$. 
$u$s and $v$s are eigenvector components in particle and hole space 
respectively.
To calculate the tunneling conductance, we must attach leads to the
Rashba nanowire. We assume the leads to be
translationally invariant semi-infinite normal leads.
The lead Hamiltonian is given by
\begin{align}
 H_{\mathrm{lead}} &= \left(-\frac{1}{2m^*}\partial_x^2 + i\alpha_R\sigma_y \partial_x - 
 \mu_{lead}
  \right)
 \tau_z  + \mu_0 B_{lead}\sigma_x. 
\label{Hlead}
\end{align}
The above lead Hamiltonian is discretized as
\begin{align}
H_{lead}^{dis} &= \sum_{n}
[-t \left(|n+1\rangle\langle n|
 + \mathrm{H.c.}\right)\tau_z \nonumber \\
&+ i\alpha \left(|n+1\rangle\langle n|
- \mathrm{H.c.}\right)\sigma_y\tau_z  \nonumber \\
&+ (2t-\mu_{lead} )|n\rangle\langle n|
 \tau_z + \mu_0B_{lead}|n\rangle\langle n|\sigma_x].
\label{Hlead1} 
\end{align}

Following the Delft experiment~\cite{Mou},
a finite applied magnetic field $B_{lead}$ 
is assumed to exist so as to have two non-
degenerate 
conducting channels because of the spin splitting induced by $B_{lead}$.  
Having a finite magnetic field in the lead also helps us to avoid the numerical challenge 
to identify and separate various channels to compute the S-matrix.
We emphasize, however, that our keeping a finite $B_{lead}$ 
is actually consistent with the experimental situation 
(and not just a matter of computational convenience).

The potential barrier defining the NS junction 
at the lead-nanowire interface (see Fig.~\ref{cartoon}) is simulated 
by  modulating the hopping
amplitude $t'$ between the nanowire and the lead. 
For higher(lower) tunnel barrier, the hopping amplitude $t'$
is lower(higher). The new system Hamiltonian
$H_{sys}^{dis} \longrightarrow H_{sys}^{'dis}$ has the form,
\begin{align}
 H_{sys}^{'dis} &= \sum_{\pmb{n}=2}^{\pmb{N}}
 [-t \left(|n+1\rangle\langle n|
 + \mathrm{H.c.}\right)\tau_z  \nonumber \\
 &+ i\alpha \left(|n+1\rangle\langle n|
 - \mathrm{H.c.}\right)\sigma_y\tau_z \nonumber \\
 &+ (-\mu +2t)|n\rangle\langle n|\tau_z \nonumber \\
 &+ V_Z|n\rangle\langle n|\sigma_x
 + \Delta_0|n\rangle\langle n|\tau_x]  \nonumber \\
 &- \left(t'|2\rangle\langle 1| + 
 \mathrm{H.c.}\right)\tau_z 
 +i\alpha '\left(|2\rangle\langle 1| - 
 \mathrm{H.c.}\right)\sigma_y\tau_z \nonumber \\
 &+ (2t-\mu_{lead}) |1\rangle\langle 1| 
 \tau_z 
 +\mu_0B_{lead}|1\rangle\langle 1|
 \sigma_x.
\label{Hbr}
\end{align}
In this setup, $t'$ and $\alpha '$ correspond to hopping and spin-orbit 
coupling between the lead and the nanowire, respectively.
$t'<<t$ would correspond to a high tunnel barrier or
weak lead-nanowire coupling. When $t'\sim t$, the tunnel barrier
is low or equivalently, the lead-nanowire coupling is strong (i.e. the barrier is almost
transparent).
The lead-nanowire tunneling $t'$ introduces a 
broadening ($\Gamma_L$) to be discussed later in the paper (c.f. Eq.~\eqref{Gamma_L}).
A strongly coupled (i.e. large $t'$) 
lead-nanowire system will have strongly broadened conductance peaks, whereas a
weakly coupled lead-nanowire system will have weakly broadened sharp peaks. 
Narrow resonances appearing from states that are weakly coupled to the lead
(as a result of being localized far away from the end) 
are removed by broadening the energy eigenstates
by introducing an on-site imaginary term in the Hamiltonian,
i.e., $H_{sys}^{'dis} \longrightarrow H_{sys}^{'dis}+
b$ where,

\begin{align}
 b &= \sum_{\pmb{n}=2}^{\pmb{N}}
 (-J i)|n\rangle\langle n|\mathbbm{1}
\label{Br}.
\end{align}

Here $J$ is the parameter controlling the intrinsic 
broadening, $\Gamma$, in the conductance profile.
The two are related by, $\Gamma = 2J$.
We note that this intrinsic broadening is again incorporated in 
the theory to be consistent with the experimental situation 
(and not just for computational efficacy) since the measured 
tunneling conductance spectra do not reflect 
sharp resonant structures even at the lowest temperatures.
Obviously, an environment-induced dissipative broadening 
(parametrized by $\Gamma$ in our theory) plays a role in the experiment.  
We emphasize that broadening plays a key role in our theory
converting the topological quantum phase transition into a crossover and providing a 
visibility for the braiding measurements.


LDOS is calculated by numerically diagonalizing
the system Hamiltonian. Throughout all our calculations, the
following set of parameters (unless specified otherwise) is used:  
$\alpha = 1.79\mathrm{K}$,
$\mu = 0\mathrm{K}$, $t=12.5\mathrm{K}$, $\Delta_0 = 3\mathrm{K}$, $L=1.5\,\mu$m, $a=54$ nm. For reasons 
motivating the choice of the parameter set, we refer the reader
to Ref.~\cite{SauSarma}. 
We believe these parameters to be a reasonably realistic 
description of the experimental situation in 
Ref.~\cite{Mou}, at least at a qualitative level.
As discussed in Appendix A, the conductance and TV are calculated from the 
scattering matrix that is obtained using 'KWANT'---a
quantum transport and simulations package in Python
 developed principally by
Groth \textit{et. al.}~\cite{groth2014kwant}.

\section{Results: Conductance and topological visibility}

\subsection{Role of broadening versus splitting}
\begin{figure}
 \begin{center}
\includegraphics[height=0.8\columnwidth,width=\columnwidth]{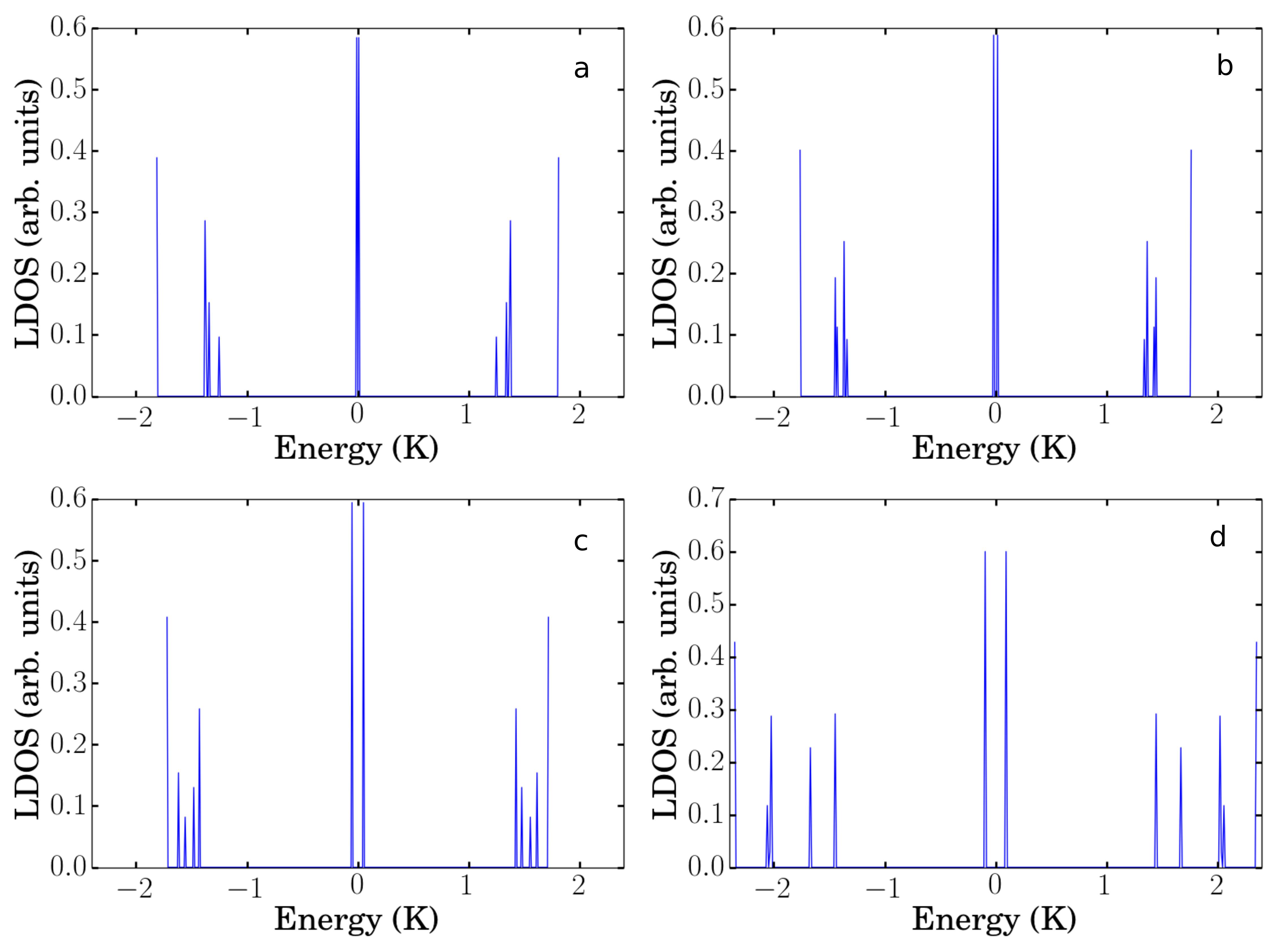}
\end{center}
\caption[]{(Color Online) LDOS for clean nanowire with 
$L = 1.5\mu m$ and Zeeman field strengths (a-d), $V_Z$ = 4.2, 4.3, 4.5 and 5.0 K.
The corresponding Majorana splitting
(a-d) are $\delta$ = 0.012, 0.036, 0.094 and 0.18 K 
respectively clearly vary strongly with $V_Z$.
}
\label{fixedDeltaLDOS}
\end{figure}

While the TV and conductance will be determined by all the microscopic 
parameters discussed in the last section, we now argue that the qualitative behavior can be 
understood in terms of a few effective parameters, which in turn are determined 
by the full set of microscopic parameters in a simple way. 
For example, as seen from the calculated 
local density of states plotted in  Fig.~\ref{fixedDeltaLDOS}, 
one of the relevant scales that affects the topological properties, 
the splitting of the MZMs ($\delta$), 
is relatively independent of the other scales such as lead coupling, 
but sensitively determined by small variations in the microscopic 
Zeeman field $V_Z$ in an oscillatory fashion~\cite{Meng,Meng1,Das1,Rai}.  
We note that $\delta$ is a key parameter determining the topological 
content of the system in the sense that when this quantity is (exponentially) 
small, the system is by definition non-Abelian, whereas by contrast, 
when $\delta$ is comparable to the 
superconducting energy gap, the system is manifestly not topological.

The topological properties of a one dimensional superconductor 
such as a semiconductor nanowire 
crucially depend on the various relevant sources of broadening, 
such as the lead coupling and inelastic 
scattering, of the quasiparticle excitations. The width of the 
ZBCP, which is a key signature of topological superconductivity, 
depends on the broadening, $\Gamma_L$, which is controlled by tuning 
the lead tunneling $t'$ discussed in the
previous section. Furthermore, the 
TV, $Q$,~\cite{Akh}, which characterizes the topology of nanowires 
with open boundary conditions, is necessarily non-topological (i.e. $Q=1$)~\cite{Akh}
because any calculation of TV in the presence of finite 
$\delta$ (which must always be true in any finite wire)  
and no broadening must necessarily give Q=1 (i.e. a non-topological 
trivial system) since the MZM is not 
located precisely at zero energy for any finite length wire!
Typically, 
this is circumvented by computing the TI at an energy arbitrarily shifted 
slightly away from zero by the splitting 
of the MZM, $\delta$. A similar behavior is noticed~\cite{lin2012zero} in the low-bias conductance $G(V)=dI/dV$, 
which  characterizes MZMs through a quantized 
value $G(V\rightarrow 0)=G_0=2e^2/h$~\cite{Sen,flensberg2010tunneling,law2009majorana,wimmer2011quantum}. 
For a finite system,
the conductance $G(V\gtrsim \delta)$ approaches the quantized value $G(\Gamma_L\gg V\gtrsim \delta)\rightarrow G_0$. On the other hand, as $V$ truly approaches zero 
(i.e. $|V|\ll \delta$), the conductance in the tunneling 
limit approaches zero~\cite{lin2012zero}, giving a vanishing ZBCP 
(since the Majorana is not located precisely at zero energy in a finite length wire).

\begin{figure}
 \begin{center}
\includegraphics[height=0.8\columnwidth,width=\columnwidth]{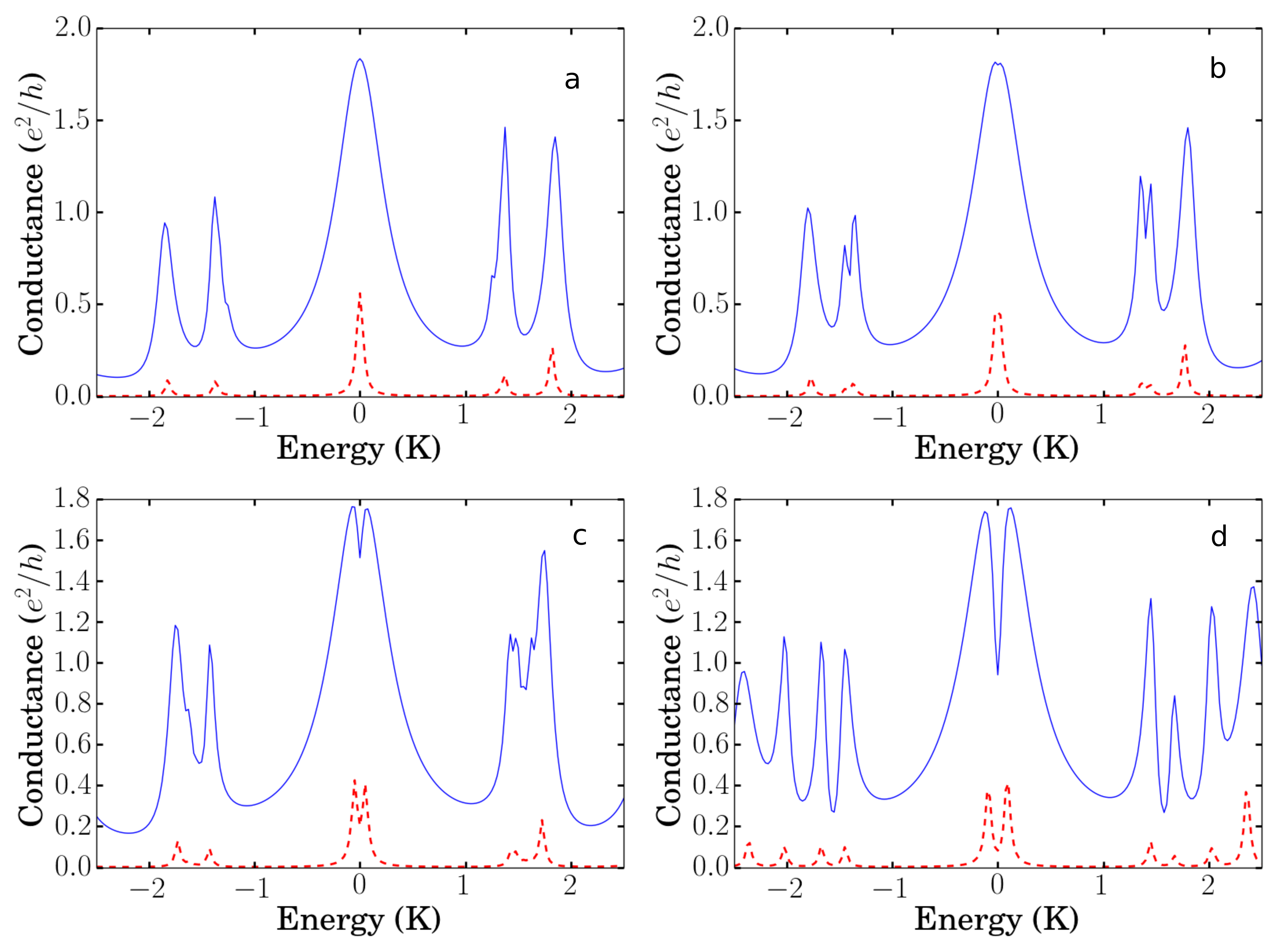}
\end{center}
\caption[]{(Color Online) Conductance plot 
corresponding
to the LDOS splittings for $\Gamma_L$/$\Gamma$ = 10 
(blue solid curve) and $\Gamma_L$/$\Gamma$ = 0.25 (red dashed curve).
The parameter 
$\delta/\Gamma $ (a-d) = 0.27 , 0.71 , 1.74  and 
3.27  respectively. 
The TV ($Q$) values (a-d) are 
(-0.80,-0.75,-0.46,0.12) and (0.44,0.58,0.82,0.93) 
for  
blue solid curve  and 
red dashed curve, respectively.
The conductance peaks split for large $\delta/\Gamma$ 
and the conductance decreases for small $\Gamma_L/\Gamma$.
}
\label{fig3}
\end{figure}

Therefore, both the TV ($Q$) and the zero-bias conductance ($G$) cannot be evaluated strictly at 
zero-energy for a finite wire to determine the topological phase of the wire. In this work, motivated by the 
goal of understanding finite rate dynamical processes such as braiding, we avoid this problem 
by introducing an intrinsic quasiparticle decay rate (i.e. dissipative broadening), 
$\Gamma$, 
which we believe to be the realistic experimental situation. 
The broadening $\Gamma$ is controlled in our calculations by choosing 
the parameter $J$ discussed in Sec. II.
The introduction of such a scale allows us to discuss the conductance ($G$) and TV 
($Q$) in a meaningful way at exactly zero energy. 
The intrinsic decay rate, $\Gamma$, apart from representing the 
uncertainty in energy resulting from the finite braiding rate, likely also exists in semiconductor wires from interactions and 
phonons (and unknown dissipative coupling to 
the environment invariably present in all solid state systems).  
Moreover, since the conductance and the TV are determined by 
the scattering properties of electrons from an external lead, 
the coupling to the lead, which is parametrized by the 
broadening $\Gamma_L$, quantitatively affects these topological properties. 
Finally, the superconducting gap $\Delta$ that protects  
topological properties themselves must play a role in determining the topological properties. In the following subsections, 
we will study the inter-dependence of the conductance and the TV on these 
energy-scales namely $\delta,\Gamma,\Gamma_L$ and $\Delta$.
We emphasize that the problem is highly complex because these are four 
completely independent energy scales (and in real experimental systems 
there are at least two additional energy scales associated with 
finite temperature and disorder neglected in the current work).

\subsection{Topological phase}
We start by discussing the zero-bias conductance and TV deep in the topological phase where 
the intrinsic quasiparticle broadening $\Gamma$ is
much smaller than the topological gap $\Delta\gg \Gamma$ 
so that the gap $\Delta$ is well-defined. We choose  the nanowire to be  sufficiently long, in this subsection, 
so that the Majorana splitting $\delta$ 
and the broadening of the MZMs from the lead are much smaller than the gap (i.e. $\delta,\Gamma_L\ll\Delta$).

Since the topological gap $\Delta$ is much larger than the parameters 
relevant to the MZMs namely, the splitting $\delta$, 
the broadening of 
the MZM due to coupling to the lead, $\Gamma_L$, and the (intrinsic)
broadening of the far end MZM (away from the lead), $\Gamma$, both the 
zero-bias conductance $G(V=0)$ and the TV $Q$, is a function only of $\delta,\Gamma_L$ and $\Gamma$. 
Note that the broadening of the MZM at the far end is the same as the intrinsic quasiparticle broadening $\Gamma$, since it is not 
coupled to the lead.
Since the absolute energy scale cannot matter, 
the conductance $G(V=0)$ and the TV, $Q$, 
can be studied as a function of 
dimensionless parameters $\Gamma_L/\Gamma$ and $\delta/\Gamma$
(in this large $\Delta$ limit).

Consider first the limit where $\delta/\Gamma \gg 1$, i.e., the broadening is much 
smaller than the Majorana 
splitting. As seen from the conductance plot in Fig.~\ref{fig3}(c,d) (red dashed curve), if the lead coupling also weak 
i.e. $\Gamma_L\ll \delta$, the conductance profile $G(V)$ shows a 
pair of resonances at energies $E=\pm \delta/2$ with broadening of order $(\Gamma+\Gamma_L)$.
The height of these peaks would be substantially below the quantized value. 
As seen from the solid blue curve in Fig.~\ref{fig3}(c,d) 
and consistent with previous work~\cite{lin2012zero}, increasing the lead coupling 
 so that $\Gamma_L\gg\delta$, increases the height of the zero energy peak so as to approach 
the quantized value $G(V\sim 0)\sim G_0$. 
However the splitting $\delta$ now appears as a dip in the conductance which reduces the conductance $G(V=0)$ 
at strictly zero-bias. Thus, the zero-bias conductance $G(V=0)$ is suppressed from the quantized value, and 
as expected from the connection between conductance and TV~\cite{wimmer2011quantum} 
, we find the TV $Q$ to be non-topological (i.e. positive in this parameter regime).

 The conductance  $G(V)$ in the opposite limit, where
$\delta/\Gamma \ll 1$, is shown in Fig.~\ref{fig3}(a,b) and shows an 
unsplit ZBCP. The conductance in the $\Gamma_L\gg \Gamma$ (blue curve) shows a nearly 
quantized conductance, while the conductance is suppressed in the opposite limit.
However, this limit (i.e. $\Gamma_L\ll \Gamma$) (red dashed curve) still shows a ZBCP, albeit 
substantially smaller than the quantized value even though the corresponding TV is non-topological. 
On a technical note,  varying the Zeeman field between the different panels in Fig.~\ref{fig3} 
changes $\Delta$.  To mitigate any parametric dependence of the 
calculated ZBCP and TV on 
$\Delta$, in this subsection
the broadening $\Gamma$ is adjusted
in each case to hold $\Delta/\Gamma = 52$ fixed 
(remembering that the gap $\Delta$ depends on the Zeeman field).
The lead broadening $\Gamma_L$ is varied through varying $t'$ 
(see Eq.~\eqref{Gamma_L} below) to keep the ratio $\Gamma_L/\Gamma$ fixed.

\begin{figure}
 \begin{center}
\includegraphics[height=0.8\columnwidth,width=\columnwidth]{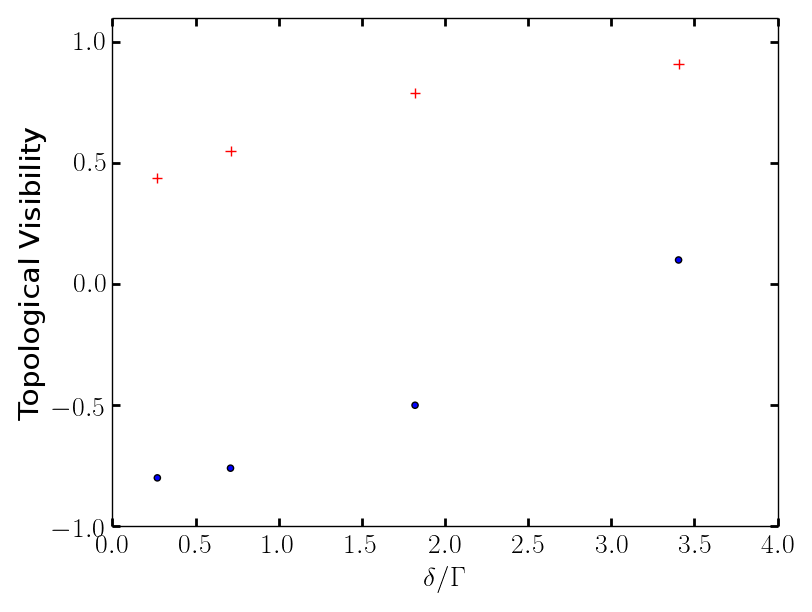}
\end{center}
\caption[]{(Color Online) TV  for $\delta/\Gamma $ values corresponding
to Fig.~\ref{fig3} for coupling parameter $\Gamma_L/\Gamma=10$   (blue dots) 
and $\Gamma_L/\Gamma=0.25$  (red plus). 
The TV is an increasing function of $\delta/\Gamma $, i.e., the system tends to 
become non-topological  as $\delta/\Gamma$ increases.
}
\label{fixedDeltaTI}
\end{figure}

\begin{figure}
 \begin{center}
\includegraphics[width = \columnwidth]{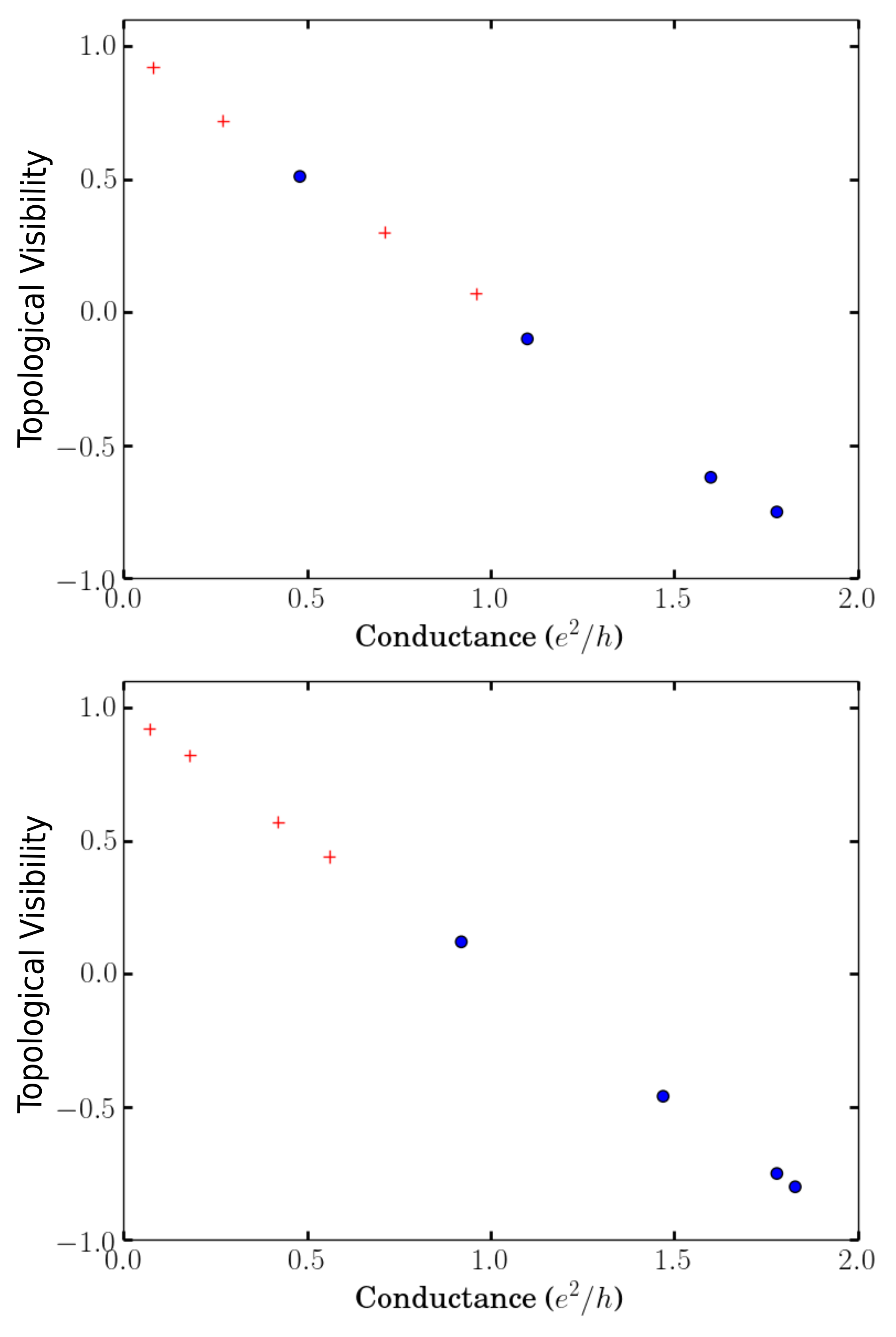}
\end{center}
\caption[]{(Color Online) (Top) Plot of the TV as a function of ZBCP for 
$\delta/\Gamma = 0.16$ (blue dot) and 
$\delta/\Gamma = 2.43$  (red plus). 
Conductance is varied by varying
$\Gamma_L/\Gamma$. (Bottom) 
TV vs ZBCP for 
$\Gamma_L/\Gamma = 10$ (blue dot) and 
$\Gamma_L/\Gamma = 0.25$ (red plus).  
Conductance is varied by varying
$\delta/\Gamma$.
The TV is a decreasing function of the ZBCP.
}
\label{TIvsCond}
\end{figure}

The TV is strongly affected by the 
 splitting of the MZMs $\delta$ relative to the broadening $\Gamma$.
In Fig.~\ref{fixedDeltaTI}, we find that the  TV is an  increasing function of 
$\delta/\Gamma$. The nanowire effectively becomes non-topological 
if the MZM splitting $\delta$ exceeds the 
broadening $\Gamma$, even when the wire parameters and the strong lead coupling $\Gamma_L$ favor the topologically 
non-trivial phase.  Furthermore, consistent with the conclusion in Fig.~\ref{fig3}, 
the small values of $\Gamma_L/\Gamma$ lead to non-topological values for the TV.

The combination of Figs.~\ref{fig3},~\ref{fixedDeltaTI} suggests a correlation between the 
presence of a quantized ZBCP and a topologically non-trivial value of the TV close to $Q=-1$.
This correlation between TV and conductance suggested by Figs.~\ref{fig3},\ref{fixedDeltaTI} 
is made explicit in Fig.~\ref{TIvsCond}. 
We find that TV is a decreasing function of the ZBCP value. 
The TV approaches -1(+1) as ZBCP
approaches $2e^2/h$($0$). Note that the decreasing behavior of TV with increasing
ZBCP is independent of the tuning parameter chosen to vary the zero bias conductance,
evidenced by the fact that both top and bottom plots in Fig.~\ref{TIvsCond} manifest 
a decreasing behavior for the TV as a function of ZBCP regardless of whether 
$\Gamma_L/\Gamma$ (top subfigure Fig.~\ref{TIvsCond}) or $\delta/\Gamma$
(bottom subfigure Fig.~\ref{TIvsCond})
is tuned to vary ZBCP.

\subsection{Topological phase transition}
Let us now consider the behavior of conductance and TV as 
we approach the TPT by tuning the Zeeman field $V_Z$.
In this case, when  the intrinsic broadening $\Gamma$ and the lead-induced 
broadening $\Gamma_L$ are small, sufficiently close to the phase transition, 
the topological gap $\Delta$ will become smaller than $\Delta\ll\Gamma$
(since at the TPT, the gap must vanish).
Therefore, for infinite length systems, the ratio $\Delta/\Gamma$ can be used to 
determine the distance to the quantum critical point.
For conventional quantum critical points~\cite{sachdev2011quantum}, 
there are two dimensionless 
parameters  
that characterize the distance to a quantum critical point, which are  $L/\xi$ and $\Delta/T$ 
characterizing spatial and imaginary time correlations in the system. 
Here $\xi$ is the coherence length of the system, $\Delta$ is an energy
scale, $T$ is the temperature 
and $L$ is the length of the system. In our discussion, $\Gamma$ is analogous to temperature $T$ in the 
quantum critical phase (although we are actually at $T=0$ throughout).
Since $\Gamma$ is always finite in our system, 
the TPT is always a crossover even at zero temperature!
The fact that our calculated TV value in Figs.~\ref{fixedDeltaTI} and~\ref{TIvsCond} 
is 
continuous between $Q=+1$ (trivial phase) and $Q = -1$ (topological phase) 
is a clear indication that the presence of dissipative broadening in the 
theory (and the associated non-unitarity) has rendered the TPT into a 
crossover with $Q > (<) 0$ defining the 
non-topological (topological) phase with finite visibility.
The presence of dissipation makes some additional changes to the topological transition that 
we mention in passing. Traditionally in disordered systems the topological transition is often 
accompanied by a Griffiths like phase populated by weakly split low-energy Majorana modes~\cite{Dam}.
The presence of dissipation could change some of these weakly split Majorana modes into poles of 
the now non-unitary S-matrix with exactly zero energy but different imaginary parts~\cite{pikulin2012weakanti}. 
Such physics, which is exactly included in our theory, 
would alter the nature of the low-energy density of states near the transition. 

The relationship between $\Gamma_L$ and $\Delta$ is not straightforward 
because as the system approaches the TPT, the bound 
states become delocalized away from the lead due to the diverging coherence length $\xi$. 
In the limit of small lead-tunneling, $t'$, the broadening $\Gamma_L$ induced by the 
lead is related to the imaginary part of the lead self-energy~\cite{datta1997electronic}
and can be written as 
\begin{align}
 \Gamma_L \sim t'^2 |\psi(0)|^2 \label{Gamma_L},
\end{align}
where $\psi(0)$ is the value of the nanowire wavefunction at the lead-nanowire
NS contact at the given tunneling energy. 
The localized Majorana wavefunction 
can be approximated by,
\begin{align}
 \psi(x) \approx \frac{1}{\sqrt{\xi}}e^{-x/\xi},
\end{align}
where $\xi$ is the superconducting coherence length.
This implies,
\begin{align}
 \Gamma_L \sim t'^2 \Delta.
 \label{Gamma_L1}
\end{align}
Therefore, in the vicinity TPT, $\Delta/\Gamma_L\propto t'^2$ (with the 
proportionality factor related to the normal phase 
density of states) approaches a constant and can be used as a parameter 
to characterize the TPT.
Note that although $\Gamma_L$ is in some sense proportional 
to the gap $\Delta$, the two quantities are still independent parameters of the theory 
by virtue of the lead tunneling matrix element $t'$.

\begin{figure}
 \begin{center}
\includegraphics[width = \columnwidth]{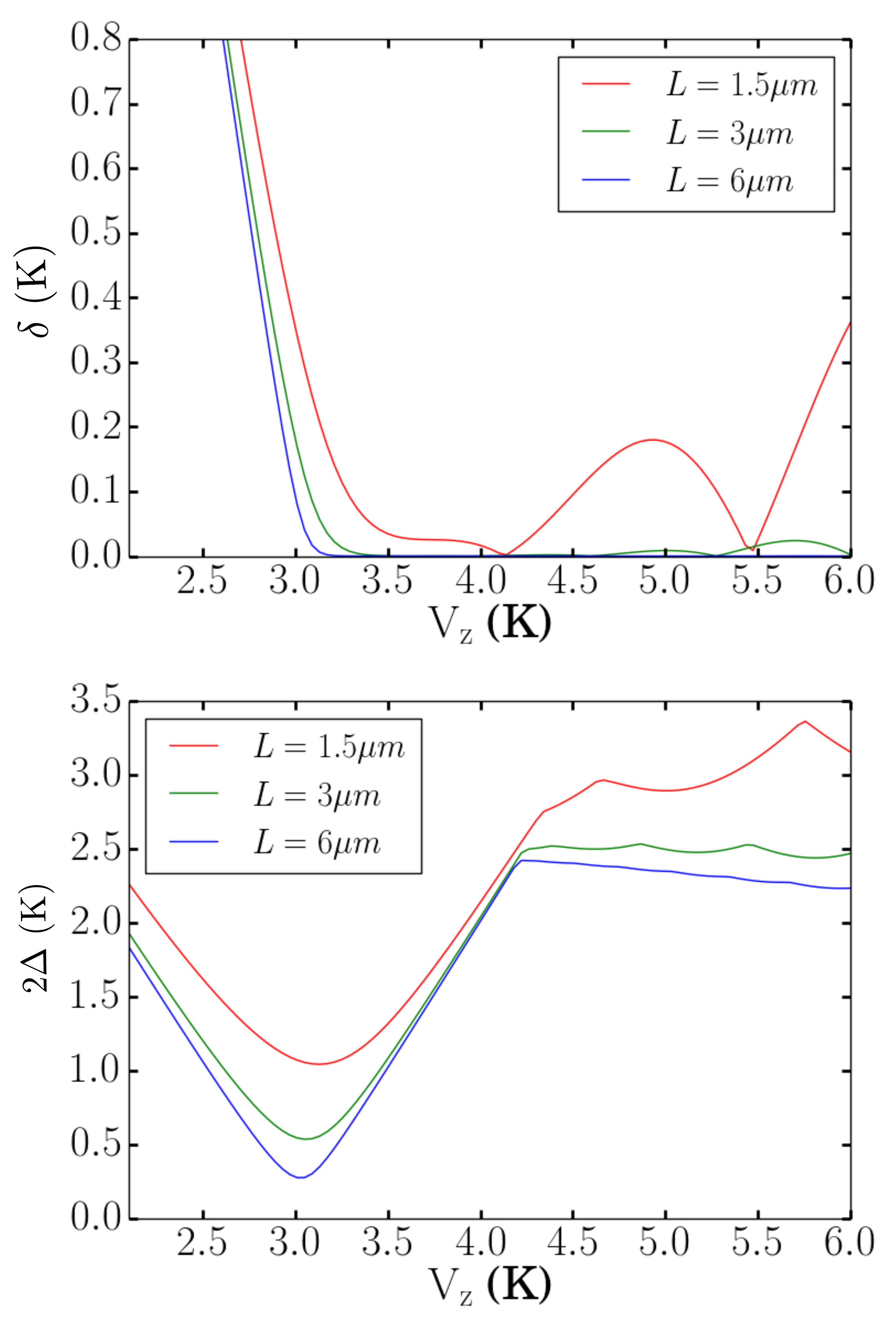}
\end{center}
\caption[]{(Color Online) Plot of Lowest Andreev bound state energy (top)
and bulk quasiparticle energy gap (bottom)
as a function of Zeeman field strength for different physical lengths
of Majorana nanowire. The bulk TPT is at $V_Z = 3\mathrm{K}$. In the topological
phase, ($V_Z > 3\mathrm{K}$),
lowest Andreev bound state energy is the Majorana splitting.
}
\label{majoranasplitting}
\end{figure}

\begin{figure}
 \begin{center}
\includegraphics[width = \columnwidth]{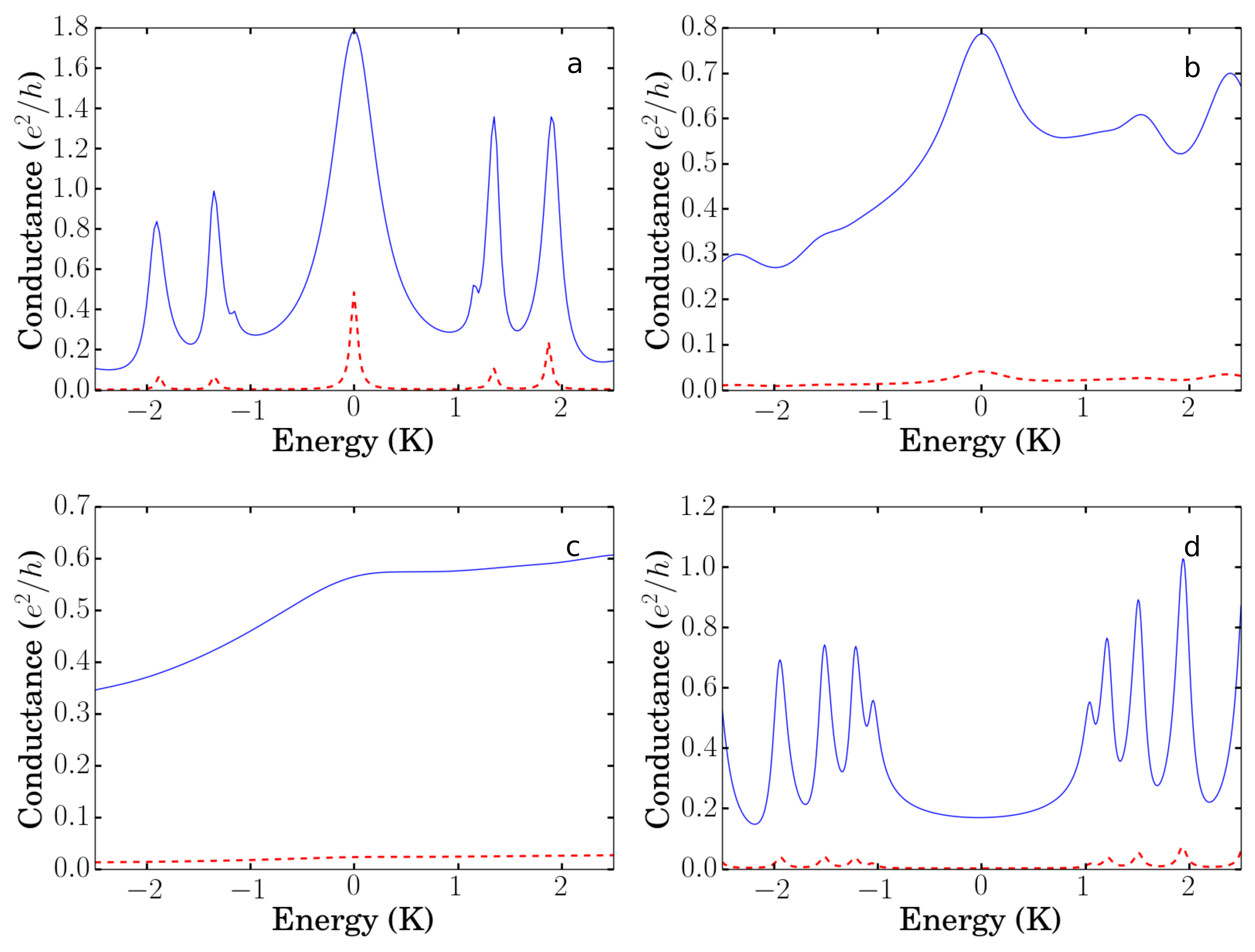}
\end{center}
\caption[]{(Color Online) 
Conductance plot 
for $\Gamma_L$/$\Delta$ = 0.20
(blue solid curve)  
and $\Gamma_L$/$\Delta$ = 0.005 (red dashed curve).
Broadening $\Gamma$ is chosen so that $\delta/\Gamma $= 0.16 
is held fixed for
all panels with
$\Delta/\Gamma $ (a-d) being 19.33, 0.90, 0.28, -12.1, 
respectively.
The TV ($Q$) values (a-d) are 
(-0.75,0.34,0.56,0.98) and (0.51,0.95,0.97,1.0) 
for 
blue solid curve  and 
red dashed curve, respectively.}
\label{fixeddeltaCond}
\end{figure}

\begin{figure}
 \begin{center}
\includegraphics[width = \columnwidth]{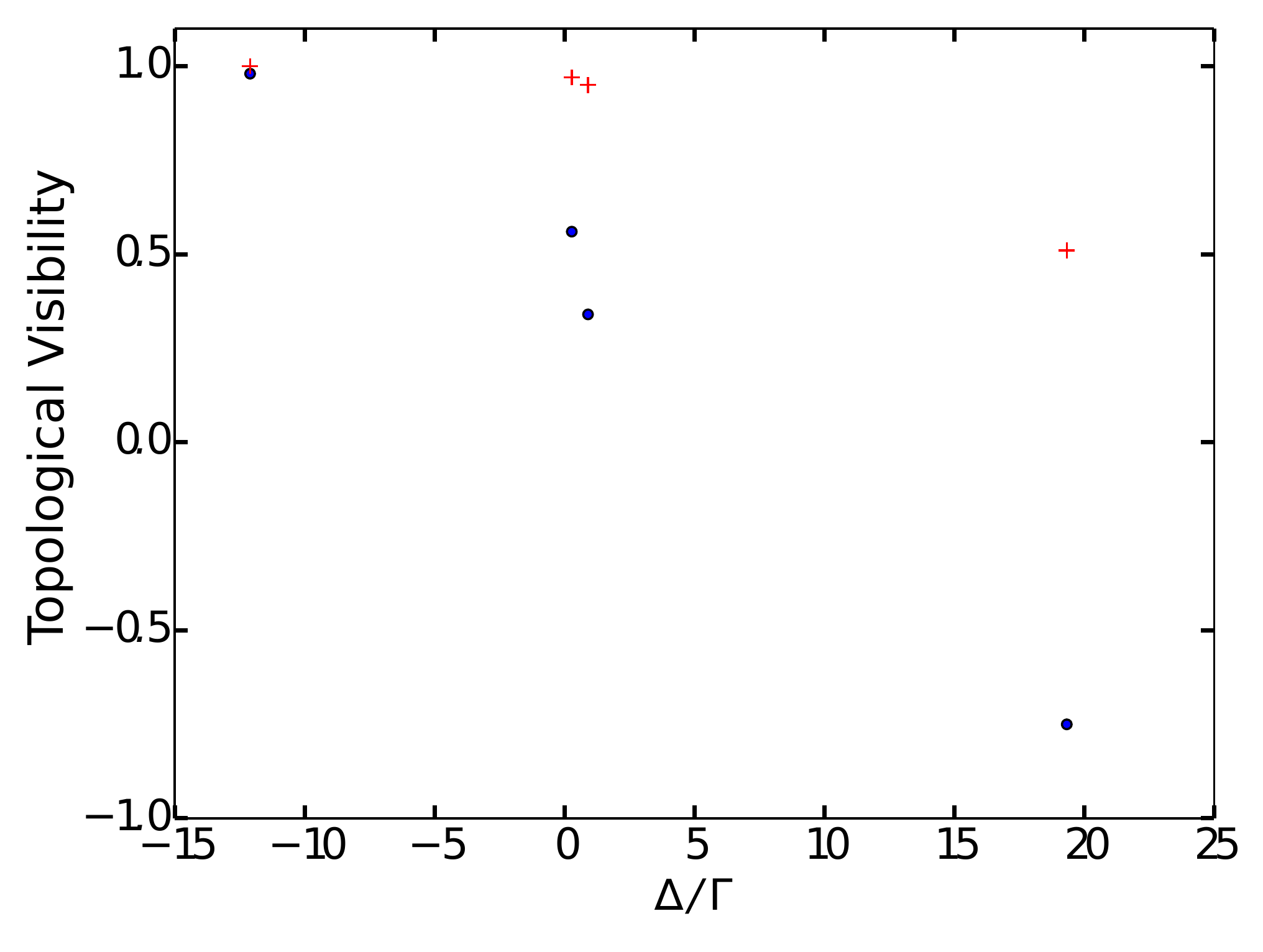}
\end{center}
\caption[]{(Color Online) TV 
for $\Delta/\Gamma $ values corresponding
to Fig.~\ref{fixeddeltaCond} for 
coupling parameters near TPT with $\Gamma_L$/$\Delta$ = 0.20 (blue dots) and 0.005 (red plus), respectively. 
$\delta/\Gamma = 0.16$ is held fixed. 
The TV is an decreasing function of $\Delta/\Gamma$ with 
$Q \longrightarrow 1$ as $\Delta/\Gamma \longrightarrow 0$, i.e., the system
tends to become non-topological (topological) as the system tends to small (large)
topological gap and the TV tends to +1 as the gap completely closes (system approaches
TPT).
}
\label{fixeddeltaTI}
\end{figure}

As seen in Fig.~\ref{majoranasplitting}, the TPT is approached by 
tuning the Zeeman field $V_Z$, 
which leads to the variation of both the Majorana splitting $\delta$ (lowest Andreev bound state energy) in the upper panel and the
bulk gap $\Delta$ (next highest Andreev boundstate energy) in the lower panel.
The minimum in the gap $\Delta$ occurs at $V_Z= 3\mathrm{K}$ 
indicating a transition at this value of the Zeeman potential. 
For a finite system, the minimum gap is determined by the length of the system $L$.
In the case where the wires are  shorter than the dephasing length 
$l_\phi\sim v/\Gamma$ (for the chosen $\Gamma$), where $v$ is the Fermi velocity of the 
system,
the MZMs split before entering the TPT region $\Delta\lesssim\Gamma$. 
As a result, the system enters a non-topological phase with a TV close to $Q=1$ 
similar to the $\delta\gtrsim\Gamma$ case discussed in the last subsection. 
Therefore 
in this section we focus on a broadening $\Gamma$ that is larger than the finite size gap i.e. $\Gamma\gtrsim v/L$.

Let us now consider the conductance shown in Fig.~\ref{fixeddeltaCond} as the 
Zeeman field is varied 
towards the topological transition. Fig.~\ref{fixeddeltaCond}(a) shows a nearly quantized peak (blue solid) deep 
in the topological phase where the MZM splitting $\delta$ is also small relative to the broadening $\Gamma$. The corresponding TV is also seen to be nearly topological in  Fig.~\ref{fixeddeltaTI} as expected.
 As the Zeeman field is decreased, the height of the ZBCP (above the background) 
 decreases as one approaches the topological 
transition where $\Delta/\Gamma\rightarrow \delta/\Gamma$ becomes small 
in Fig.~\ref{fixeddeltaCond}(c).
However, it should be noted that the peak remains unsplit in contrast to the short wire case with $L\lesssim l_\phi$.
Despite the presence of a small zero bias peak in Figs.~\ref{fixeddeltaCond}(b) and \ref{fixeddeltaCond}(c), 
the corresponding TV values in 
Fig.~\ref{fixeddeltaTI} are positive (non-topological). 
This is consistent with Figs.~\ref{fig3} and~\ref{fixedDeltaTI}
from the previous subsection where a small coupling $\Gamma_L\ll\Gamma$ led to small non-topological ZBCP.
Finally, as one crosses over to the non-topological regime,
a non-topological gap appears in the 
conductance. As mentioned before, the TPT is parameterized by $\Gamma_L/\Delta$, which remains relatively 
constant near the phase transition. 
The red dashed plots in Fig.~\ref{fixeddeltaCond} show that the conductance is 
systematically suppressed in the regime of small $\Gamma_L/\Delta$. 
The corresponding TVs are seen to be positive (non-topological)
in Fig.~\ref{fixeddeltaTI}.

Before concluding this section, we comment on an obvious point which
might confuse a non-alert reader.  One may think that the TV 
can have only unit magnitude with $Q$=-1 (+1) characterizing the
topological (trivial) phase.  This is indeed so in the infinite system as 
originally introduced by Kitaev.  But our finite system must have a 
broadening (otherwise the TV calculated at zero energy is always +1 because 
of Majorana splitting), and this broadening allows the TV (i.e. $Q$) to be 
a continuous function of system parameters going from +1 deep in the trivial 
phase to -1 deep in the topological phase.  This continuous evolution of $Q$ 
between +1 and -1 is the finite system crossover transition whereas the corresponding 
infinite system would have a sharp transition from +1 to -1 precisely at the 
TPT (with the ZBCP value changing from zero to $2e^2/h$ sharply at the TPT too).  
The new idea in the current work is to connect this crossover transition to 
braiding experiments with the claim that our finding a value of $Q<0$ corresponds
to a topological phase with the visibility of the braiding measurements being 
large (small) depending on whether the magnitude of $Q$ is close to unity (zero).
We believe that our finding a negative (positive) value of
$Q$ corresponds to the corresponding 
braiding experiment manifesting (not manifesting) non-Abelian statistics.

\section{Braiding and tunneling conductance}

\begin{figure}
 \begin{center}
\includegraphics[width=\columnwidth]{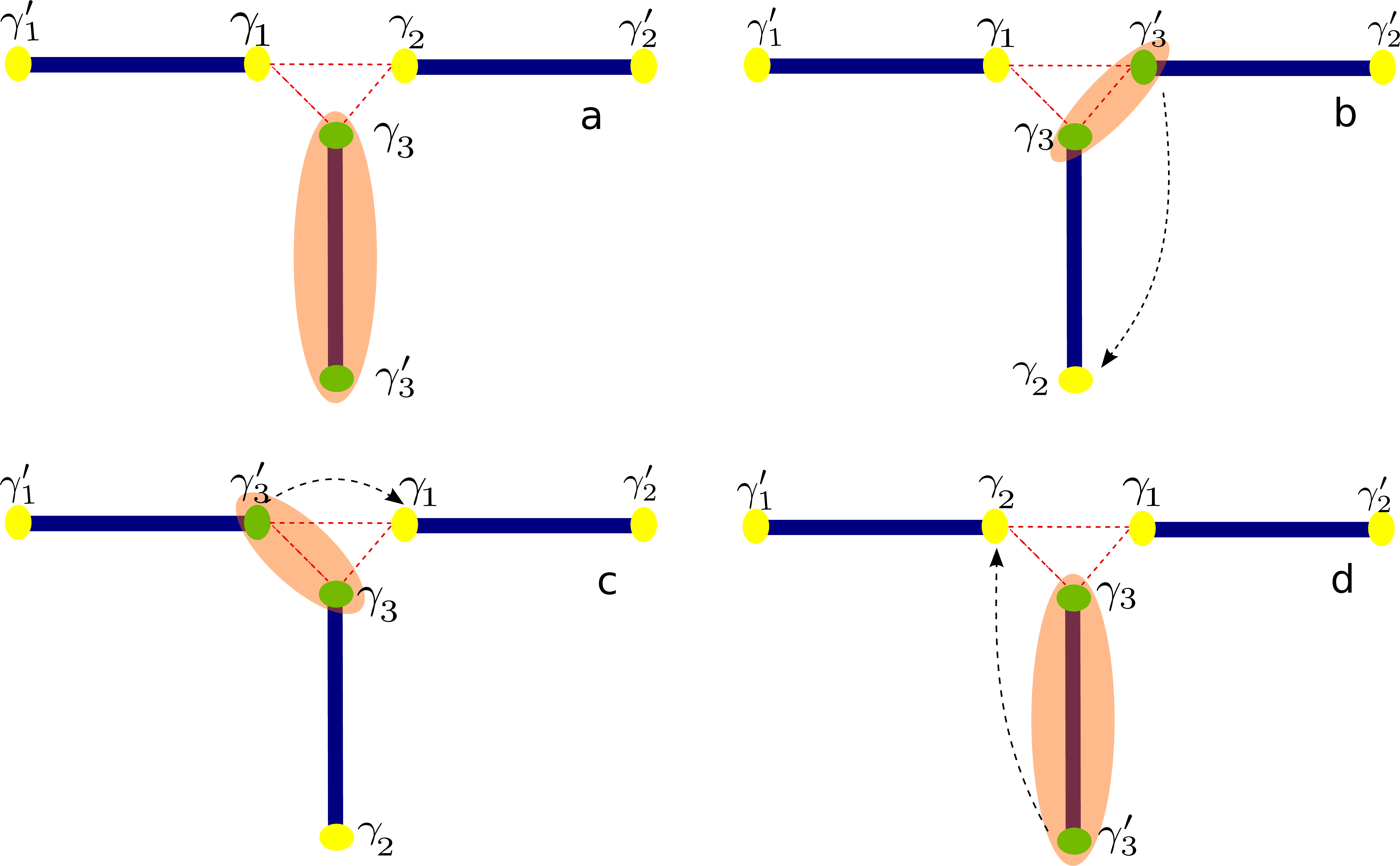}
\end{center}
\caption{(Color Online)
A schematic diagram for braiding a pair of Majoranas using a tri-junction. The 
system is initialized (a) such that $\gamma_1'$, $\gamma_1$,
$\gamma_2'$ and $\gamma_2$ are four localized Majorana modes 
and ($\gamma_3'$,$\gamma_3$)
Majorana pair is paired into a Dirac fermion.
Paired Majorana modes are depicted as green discs paired by enhanced 
wavefunction overlap over the region depicted by pink oval. 
At every move an  unpaired 
Majorana mode is moved from one position to another. 
The movement of the Majorana resulting from the move resulting 
in each configuration (b-d) is shown by a dashed arrow.
}
\label{cartoon3}
\end{figure}

In a 1D system, MZMs appear as a pair of zero energy modes (i.e. precisely at 
mid-gap assuming no 
Majorana splitting, i.e. no overlap between the two MZM wavefunctions)
localized spatially at the two wire ends. 
$N$ such pairs of localized MZMs form a $2^N$ dimensional 
degenerate zero energy subspace within
the Hilbert space of the system. A possible braiding process involving two MZMs via
a so called tri-junction~\cite{sau2011controlling} is depicted in Fig.~\ref{cartoon3}. 
Let $\hat{\gamma}_{1,2}$ and $\hat{\gamma}'_{1,2}$ represent the Majorana operators, 
associated with the localized Majorana modes $\gamma_{1,2}$ and $\gamma'_{1,2}$
depicted in the Fig.~\ref{cartoon3}a.
The ground state is initialized as a direct product state of
left and right Majorana pairs, $|P_L\rangle_1|P_R\rangle_2$,
 where $\hat{\gamma}_1$ and $\hat{\gamma}'_1$ operate on kets with subscript 1 and 
 $\hat{\gamma}_2$ and $\hat{\gamma}'_2$
 operate on kets with subscript 2. 
$P_L$, $P_R$ denote the fermionic parity  of the left 
 and right  subsystem of the initial state, respectively.
Majorana modes represented by 
$\gamma_1$ and $\gamma_2$ are braided around each other adiabatically 
via a sequence Majorana
moves involving a third Majorana pair as shown.
For example, the configuration in panel b is attained from configuration in panel a 
by adiabatically decreasing the tunneling
strength between Majorana modes $\gamma_3$ and $\gamma_3'$ and at the same time
increasing the tunneling strength between $\gamma_3$ and $\gamma_2$, which
effectively leads to $\gamma_2$ moving to the new position as
shown by the arrow in (b).
Assuming the system remains in the ground state throughout the braiding process,
the braiding is accomplished by the unitary operator,
\begin{align}
 U_{12} = \mathrm{exp}\left(\pm\frac{\pi}{4}\hat{\gamma}_1\hat{\gamma}_2\right),
\end{align}
where +/- sign in the unitary operator depends on the microscopic
details of the system~\cite{sau2011controlling}.
Considering a more complicated setup one could imagine braiding any two 
Majoranas independently of the other two. The unitary operators
affecting the braiding operation between any two Majoranas do not
form a commuting set. Hence, MZMs are said to have non-Abelian braiding
statistics which offers immense potential for 
topological quantum computation~\cite{Che,sarma2015majorana}. 

However, any realistic braiding experiment must take into account a few 
prominent departures from the idealized set of implicit assumptions made above in our
schematic description of perfect Majorana braiding. 
First, any finite system hosting MZMs
will have a finite Majorana wavefunction overlap, splitting the Majorana
modes by an energy $\delta$, away from zero energy due to the hybridization 
between the two MZM wavefunctions from the two wire ends~\cite{lin2012zero}. 
Obviously, a large overlap (as would happen in shorter nanowires or in systems 
with small superconducting gaps leading to large coherence lengths) 
would completely destroy all non-Abelian topological properties since the 
Majorana excitations in that situation are simply the 
electron-hole quasiparticle excitations of the 
superconducting nanowire
with the Majorana splitting being comparable to the superconducting energy gap.
Including this Majorana splitting in the 
formalism is an important ingredient of our theory.
Second, ``adiabatic'' braiding process takes place 
over a finite time scale $\delta t_B$ (i.e. with a finite braiding velocity), 
which is associated with the energy uncertainty of the system $\delta E_B$ by
\begin{align}
 \delta t_B \delta E_B \sim \hbar.
 \label{dtde}
\end{align}
We note that this braiding-induced energy 
uncertainty $\delta E_B$ must be much larger (smaller) than the Majorana 
splitting (superconducting gap) for the braiding operation to manifest 
any topological non-Abelian behavior.  One can loosely identify this 
energy uncertainty as 
an effective dissipation term arising from the finite velocity braiding process.
Including an energy broadening or a dissipative term is 
a key ingredient of our theory.  Such dissipation 
could arise from the energy uncertainty associated with 
braiding as discussed above, but in the specific context of 
the tunneling conductance measurements, it arises from intrinsic dissipation 
of strength $\Gamma$, 
which might be present in the experimental situation.
In the case of braiding we will refer to this intrinsic dissipation as $\Gamma'$ and use $\Gamma$
 to represent the total effective dissipation $\Gamma= \Gamma'+\delta E_B$, that also includes the energy uncertainty
 $\delta E_B$. 
 Note that $\Gamma$ is finite in the braiding process even when intrinsic 
 dissipation (i.e. $\Gamma '$) is absent since $\delta E_B$ is necessarily finite.  
 One may think that it is in principle possible to make $\delta E_B$ vanish by carrying 
 out the braiding process infinitely slowly (i.e. $\delta E_B = 0$), but this is not 
 allowed (even as a matter of principle) since the Majorana splitting is always finite
 in any finite wire, and $\delta E_B$ must always surpass Majorana splitting for the system 
 to act non-Abelian.  This is equivalent to our earlier statement that a finite wire can
 never have a true Majorana-induced zero bias conductance peak because of 
 Majorana splitting, and the presence of various energy broadening mechanisms 
 (e.g. dissipation, temperature, coupling to the leads) is essential in producing the
 ZBCP in finite wires even in the nominal topological phase. 
 Thus, braiding success and ZBCP are conceptually connected with dissipation playing a central role.

While one might argue that braiding experiments differ fundamentally 
from conductance experiments since the latter depends on $\Gamma_L$ and the
former does not, the existing braiding proposals~\cite{Jas2,sau2011QComp,clarke2011QComp,hyart2013flux,
amorim2015majorana,van2012coulomb,liang2012manipulation,romito2012manipulating,
halperin2012adiabatic,kotetes2013engineering,clarke2015bell,liu2013manipulating,
chiu2015majorana,sau2011controlling,sau2010universal} 
requires the presence of Majorana fermion tunneling in a key way. 
The Majorana tunneling enters through the tri-junctions in the Majorana 
braiding proposals. In fact, the energy gap (i.e. Majorana splitting) between
Majorana modes at the tri-junction is generated by tunneling and takes the place of the tunnel broadening
$\Gamma_L$ in the conductance experiment. Thus, inclusion of energy 
broadening $\Gamma\sim\delta E_B+\Gamma'$ (to represent finite braiding velocity
as well as any intrinsic broadening), 
tunneling broadening $\Gamma_L$, and Majorana splitting $\delta$ are 
essential  ingredients in the braiding process as much as they are in
the tunnel conductance experiments. 
The TV 
calculations discussed in the last section thus are fundamentally 
relevant to the braiding experiments as they are to the conductance experiments since
braiding is an operational way of measuring the TI 
of the system, which we have argued gets converted to TV in finite wires with dissipation.

Let us now discuss if the qualitative dependence of the braiding 
properties on the parameters $(\Gamma,\Gamma_L,\delta)$ 
is similar to their counterparts in the case of conductance.
Analogous to the tunneling 
case, the proper topological movement of MZMs (i.e. the MZMs remaining localized on the time-scale 
of the braiding operation) requires 
that the velocity-induced broadening of Majoranas satisfy
$\delta\ll \Gamma$
since any braiding must involve an actual physical movement of MZMs around each other.
Similarly $\Gamma_L$ limits the speed of braiding so that for sucessful braiding 
we require $\Gamma_L\gg \Gamma=(\delta E_B+\Gamma')$.
Furthermore, to ensure the presence of MZMs at the ends of the topological set-up, 
the finite-size Majorana splitting $\delta$ must be smaller than broadening, i.e., $\Gamma_L\gg \delta$.
We have assumed here (for the sake of this discussion without any loss of generality) 
that we are deep in the 
topological phase so that all parameters $(\Gamma_L,\Gamma,\delta)\lesssim \Delta$.
i.e., the topological gap is large, which is necessary for adiabaticity in braiding any way.

We therefore see a one-to-one correspondence between the parameters determining braiding and tunneling 
measurements with the TV showing up in both measurements as 
the key quantity determining the topological behavior of the circuit.
The braiding properties of the system might be 
characterized by $P_{\mathrm{braid}}$, which we define to be the probability of success of non-Abelian braiding.
The probability of successful non-Abelian braiding, directly related to the TV 
discussed in the last section,  is
a function
of the amount of non-universal broadening $\Gamma$ present in the braiding experiment 
(i.e. the sum of the energy uncertainty $\delta E_B$
and the intrinsic broadening due to coupling to the environment $\Gamma '$), the
Majorana splitting ($\delta$), the tunnel coupling $\Gamma_L$ and the topological gap ($\Delta$).
Furthermore, since braiding is presumably a topological property, 
we expect the probability of success of braiding to be determined by the TV
since the topological phase for the infinite system is defined by the TI.

Based on these considerations we conjecture that the success rate
of non-Abelian braiding for a given 
braiding speed in an experiment ($P_{\mathrm{braid}}$)
is related to probability of TV being $-1$ (or very close to it), i.e.,
\begin{align}
 P_{\mathrm{braid}}(\delta E_B + \Gamma ', \Gamma_L,\delta, \Delta) 
 \sim \frac{1-\langle Q(\Gamma,\Gamma_L,\delta, \Delta)\rangle}{2},
 \label{ProBraid}
\end{align}
where $\langle Q\rangle$ is the average of TV over disorder realizations for a
given disorder strength, and $\Gamma '$ is the environment-induced intrinsic
broadening in the braiding experiment. 
$\Gamma_L$ in a tunneling conductance experiment represented in the RHS of Eq.~\eqref{ProBraid} is the lead
broadening as discussed in the previous sections. However, $\Gamma_L$ appearing in the LHS of Eq.~\eqref{ProBraid} 
represents the induced tunnel gap as a result of strongly coupled adjacent Majorana modes forming a Dirac fermion 
(strong Majorana pairing regions depicted by pink ovals in Fig.~\ref{cartoon3}). The role
played by lead induced broadening for conductance experiment is now played by the energy gap induced by coupling adjacent
nanowire edge modes forming a Dirac fermion in the braiding experiment, and therefore for the sake of brevity we have chosen
to represent it with the same symbol $\Gamma_L$ on both sides of Eq.~\eqref{ProBraid}
although the $\Gamma_L$ appearing in  conductance and braiding experiments arises from different tunneling mechanisms.

From the previous section, we know that whether the average TV $\langle Q\rangle$ 
is nearly topological (i.e. a negative number with magnitude close to unity), 
which (according to our conjecture Eq.~\eqref{ProBraid})
would correspond to successful braiding,
is directly correlated with the presence of a ZBCP value
close to the topological value $G(V\sim 0)\sim G_0$.
Such a nearly quantized ZBCP, which can be tested for through existing 
experimental set-ups~\cite{Mou,Den,Das,Fin,Chu,Cha}, 
can only occur in a much smaller parameter regime $\Delta\gg \Gamma_L\gg\Gamma\gg \delta$. 
Furthermore,  temperature, which provides a fifth independent energy 
scale through the thermal energy $k_BT$ (which we take 
to be zero) must be small as well. It is only 
in this topological parameter regime that one expects braiding to be reasonably successful. We believe that this 
parameter regime can be diagnosed from the much simpler conductance quantization measurements
carried out on the same or similar samples (i.e. with similar values of $\Gamma$ and $\Gamma_L$ in both cases).

Our results (see Fig.~\ref{TIvsCond}) in section IV indicate that a  ZBCP value
around half of 
the quantized value (i.e. ZBCP $\sim e^2/h$) should be adequate to produce a 
negative TV value. 
The negative TV would correpond to the topologically non-trivial phase with a TI of -1.
Based on this we conclude that braiding experiments 
would succeed (perhaps with rather low visibility) as long as the corresponding 
ZBCP is around $e^2/h$ (or larger) in the same nanowire sample with identical system parameters.  
We believe that for systems with ZBCP much lower than $e^2/h$, the braiding experiments
are 
unlikely to succeed in manifesting a purely topological phase with a TV value of -1.
This is an important predicted experimental consequence of our theory.
We therefore believe that braiding measurements should only be attempted on 
nanowire samples with the largest possible ZBCP values, and perhaps, braiding in 
samples with ZBCP values much lower than $e^2/h$
is unlikely to manifest non-Abelian statistics (even with low visibility-- see our Fig.~\ref{TIvsCond}).

\section{Conclusion}
In this work, we ask whether a theoretical connection can be established 
between the tunneling conductance and the topological visibility  of realistic 
spin-orbit coupled semiconductor nanowires in the presence of proximity induced
superconductivity and Zeeman splitting, assuming the system parameters 
(Zeeman splitting, chemical potential, superconducting energy gap) 
to be in the topological phase so that the wire carries Majorana zero modes 
localized at the wire ends.  The question takes on special significance because 
of the putative non-Abelian braiding properties of MZMs enabling fault tolerant 
quantum computation.  In particular, direct braiding experiments, which are 
typically very hard, establishing the non-Abelian nature of MZMs have not yet 
been carried out in semiconductor nanowires although many proposals to do so 
exist in the theoretical literature.  (Such experiments do exit in the fractional
quantum Hall context for the so-called 5/2 fractional quantum Hall state, but the
results are difficult to interpret and have remained 
controversial (see Ref.~\cite{willet2013qh} and references therein). 
On the other hand, several groups have carried out tunneling conductance 
measurement in semiconductor nanowires following specific theoretical predictions 
that MZMs should manifest as zero bias conductance peaks in such experiments. 
The observation of such ZBCPs has so far been hailed as the evidence for the 
predicted existence of MZMs in these nanowires, but doubts remain about how 
topological such systems are (even if the observed ZBCP signal indeed arises from 
MZM-related physics and is not some spurious effect), particularly in view of the 
disturbing fact that the measured ZBCP values are substantially lower than the 
quantized conductance value (i.e. $2e^2/h$) expected from the perfect Andreev 
reflection by the Majorana modes.  Even assuming that the system is in the 
topological phase as far as parameter values go, serious issues arise from the finite
length of the wires, which, coupled with the expected long coherence length because 
of the small induced superconducting gap~\cite{sarma2015substrate}, 
leads to questions regarding the overlap 
between the MZMs localized at the two wire ends.  Such MZM hybridization would lead
to large Majorana energy splittings, and the MZMs would be shifted from zero-energy, 
becoming instead finite energy resonances in the gap.  If the Majorana energy 
splitting is comparable to the energy gap itself, then obviously there can be no
non-Abelian braiding statistics since these split Majorana modes are essentially 
electron-hole pairs.  The quantitative technical question now becomes whether such
approximate (or almost)- MZMs, which are split and thus shifted from zero energy, 
could still lead to non-Abelian statistics although the ZBCP associated with them 
is below the quantized conductance $2e^2/h$.  We address this question in great detail
by calculating both the tunnel conductance and the TV of the same
realistic nanowire (i.e. using exactly the same parameter values) and comparing them
carefully.
 
The TV in the ideal situation is +/- 1 with the 
negative (positive) sign corresponding to the topological (trivial) phase, 
just as the tunnel conductance in the ideal situation is $2e^2/h$ ($0$) for the 
topological (trivial) phase.  But, in real measurements, the existence of 
Majorana splitting in finite wires plus various dissipative 
broadening mechanisms invariably 
present in real systems could lead to a value of the TV with 
a magnitude less than unity, just as the ZBCP magnitude could be less than $2e^2/h$
from the same physics.  Correlating the two quantities in realistic wires would 
tell us whether braiding experiments are likely to succeed in realistic nanowires
currently being studied in various laboratories. 
One aspect regarding Majorana nanowires
is that a naive calculation of the ZBCP in the presence 
of Majorana splitting in finite wires always leads to zero conductance at zero 
energy since the finite energy split Majorana resonances have no spectral weight
at zero energy.  This of course changes in the presence of any energy broadening 
which must invariably be present in real systems. This is quite analogous to the situation 
of purely adiabatic braiding, where braiding at a rate much smaller than the Majorana splitting 
$\delta$ would also produce purely non-topological results. Our work explicitly includes 
this energy broadening effect in order to comment on real systems of experimental
interest.  We believe that our calculated TV in realistic 
systems provides  the actual visibility of future braiding measurements through 
the inclusion of broadening processes, i.e., our finding a TV
differing in magnitude from unity has one to one correspondence with the corresponding
averaged braiding experiment runs over many measurements (where the average will 
differ from unity in magnitude although each run itself will give a value of +/-1).
We note that indeed the calculated TV is negative or positive
(but always $<$1 in magnitude), depending on whether the system is approximately 
topological or trivial,  respectively.  The exact value of our calculated topological
visibility predicts the outcome of braiding experiments--- closer our results are 
to -1 more non-Abelian is the system, but any negative value for the topological
visibility could be construed as predicting the system to be in the non-Abelian 
topological phase, albeit with a low visibility if the value of the topological 
visibility is far from -1.  On the other hand, our finding of a positive topological
visibility indicates that the corresponding system is non-topological.

Our work shows that the topological quantum phase 
transition separating the trivial phase 
(a TV value of 1 and a ZBCP value of zero) from the topological 
phase (a TV value of -1 and a ZBCP value of $2e^2/h$) 
is a crossover in real systems (even at zero temperature) 
because of the presence of the broadening terms $\Gamma$, tunneling 
$\Gamma_L$ and the Majorana splitting ($\delta$).  
The inclusion of the dissipative broadening processes,
which must invariably be present in real systems, is a key 
ingredient of our theory--- in fact, without any broadening, 
the ZBCP is always zero at zero energy by virtue of the 
Majorana splitting in all finite wires.  
We find that the ZBCP 
evolves from a quantized peak deep in the topological phase into a 
much smaller peak on a large background near the transition,
quite similar to some of the experimental results~\cite{Mou,Den,Das,Fin,Chu,Cha}.
Unfortunately, the corresponding TV in this case is invariably non-topological
as long as the ZBCP value is small.
In braiding experiments
the broadening is to be interpreted as the energy uncertainly associated 
with the finite braiding time, which should be large compared 
with the Majorana splitting for braiding to succeed 
(i.e. $\Gamma>\delta$ must apply for the TV to be negative). 
For braiding to succeed, of course, one must always be deep in 
the topological gap so that the topological gap is much larger 
than the Majorana splitting ($\delta$),
and intrinsic broadening ($\Gamma$).
Furthermore, the Majorana coupling energy $\Gamma_L$ must be large enough to 
overcome the splitting $\delta$ to lead to a large conductance and also for the 
tri-junctions in a braiding protocol to lead to non-Abelian statistics.

We find that it is possible for the system to be topological 
(i.e. negative value for the TV) even when 
the corresponding zero bias conductance value is suppressed from $2e^2/h$--- 
in particular,
a factor of 2 suppression of the ZBCP would still lead to the existence of 
non-Abelian braiding statistics (with somewhat low visibility). 
On the other hand, we believe that systems with ZBCP values 
factors of 10 (or more) suppressed from $2e^2/h$ are unlikely to ever 
manifest non-Abelian statistics, and such systems are better 
considered as
non-topological systems because of the very large Majorana splitting
in spite of there being a small ZBCP peak.
Our most important qualitative conclusion is the finding that it is indeed 
possible for a finite wire with split MZMs (and a correspondingly suppressed 
ZBCP value compared with $2e^2/h$) to manifest non-Abelian braiding statistics 
with the visibility of braiding (averaged over many runs) decreasing with decreasing
value of the corresponding ZBCP.   How small the ZBCP can be and still reflect 
an underlying non-Abelian braiding statistics depends on many details 
(most importantly the ratio of the Majorana splitting energy to the topological 
gap which should typically be less than 0.2 for braiding to succeed) including 
the energy broadening in the system arising from many non-universal mechanisms.
One important conclusion following from our extensive numerical 
simulations is that braiding experiments are perhaps likely (unlikely) to succeed 
in nanowires manifesting ZBCP values at least around $e^2/h$ (much less than $e^2/h$) 
since we find that the calculated topological visibility crosses over from being 
negative to positive for the corresponding tunneling  ZBCP value crossing over 
respectively from being $>e^2/h$ to being $<e^2/h$.  
Of course, real braiding experiments
would obviously not be carried out in the tunneling configurations with leads to 
normal contacts for measuring tunneling currents, but our results indicate that 
braiding should focus on nanowires manifesting ZBCP values not much less than $e^2/h$ 
for a 
reasonable chance of any success in the observation of non-Abelian braiding statistics.
We mention that braiding experiments still involve aspects 
of tunneling (i.e. a finite $\Gamma_L$) which must arise from the finite 
Majorana tunneling at wire junctions (in contrast to NS junctions in the 
conductance measurements) necessary to make the 
MZMs go around each other in order to accomplish braiding.

Finally, we note that we have neglected finite temperature and disorder effects 
in our theory, assuming clean nanowires at zero temperature 
in order to consider the best case scenarios.  
We have assumed zero temperature for simplicity and to avoid introducing 
extra parameters, even though 
it is rather simple to introduce finite temperature effects into the conductance calculations. This is 
because finite temperature conductance of a non-interacting system can be obtained simply 
by broadening the conductance traces by a fermi function. The result of such a broadening is easy 
to surmise from the zero temperature plot. The most important effect of introducing 
temperature would be to potentially suppress the zero bias conductance peak and generally eliminate
sharp features, quite similar to the broadening $\Gamma$ that we have introduced. However, 
this is significant only if the temperature $T$ is large enough (i.e. $T\gtrsim \Gamma$). This limit 
is easy to detect in experiments since the width of the peak should correspond to temperature. Therefore, 
our results focus on the limit where temperature is low enough so as to be smaller than the width of the 
peak as in the recent experiments \cite{delftballistic}. Additionally finite temperature does not 
invalidate our conjecture regarding braiding since T must also be smaller than $\Gamma_L$ (related to the gap) 
for successful braiding. Since nanowire conductance and braiding experiments are carried out at very low 
temperatures ($\sim$20-50 mK) any way, our neglect of finite temperature is not a serious problem.
Including disorder effects is also straightforward and only increases the 
computational
time substantially (without introducing any theoretical complications),
which is why they are left out.  We emphasize that our conclusion remains 
completely unaffected by finite temperature and disorder.  
Finite temperature only reduces the visibility, thus further reducing the magnitude
of the ZBCP and the TV, without affecting the topological or not question at all.  
Thus, the braiding experiment should be performed at the lowest possible temperatures 
to maximize the visibility.  
Disorder complicates matters only because it shifts the condition for
obtaining the topological phase (i.e. the TPT point), 
but it cannot affect the basic physics at all since the induced topological 
superconductivity arises from an interplay among the s-wave superconductivity, 
spin-orbit coupling, and Zeeman splitting--- all of which are immune to disorder.  
The fact that disorder does not suppress the topological phase 
(but does shift its location on the phase diagram compared with the clean parameters)
is already well-known in the literature~\cite{adagideli2014effects}, 
and we therefore refrain from providing finite disorder results since 
this will only complicate the presentation with no additional conceptual 
or theoretical understanding. 
The situation with very strong disorder is, however,
disastrous for the manifestation of topological properties since 
the strongly disordered nanowire will manifest Griffiths phase physics 
with many MZMs localized randomly along the wire~\cite{Dam,cole2016proximity},
and this situation must 
obviously be avoided at all costs for all braiding experiments.
We have ensured numerically that all our
conclusions in this paper remain unaffected in the presence of finite 
temperature and (weak) disorder as asserted above.
Similarly,  multisubband occupancy of the nanowire 
~\cite{lutchyn2011search,stanescu2011majorana} 
does not 
change any of our conclusions either as long as an odd number 
of spin-split subbands are occupied in the system, and the 
appropriate microscopic parameters 
(i.e. $\delta$, $\Gamma$, $\Gamma_L$, $\Delta$, 
chemical potential) are all modified to take into account the 
multisubband occupancy in the nanowire.
Of course, the relative values of the various parameters may be 
modified by multi-subband occupancy, which must be incorporated in 
the theory appropriately, but the theory itself remains exactly the 
same as long as an odd number of subbands are occupied in the nanowire 
and various parameters are 
appropriately modified to reflect the multi-subband occupancy of the system. 

The new important concept introduced in this work is of 
topological visibility, which is essentially the 'nonunitary' version of the 
well-known 'topological invariant' extensively used to characterize topological 
superconductivity.  Whereas the topological invariant is a topological 
index, being +1 or -1 corresponding to trivial and topological superconductors 
respectively, the topological visibility by contrast corresponds to a 
continuous variable (varying between +1 and -1) relevant for finite systems 
where a naive computation of the topological invariant will always indicate 
a trivial phase by virtue of the Majorana energy splitting always being finite 
in finite systems. The topological visibility is a physical (and practical) 
generalization of the mathematical concept of topological invariant to realistic 
finite nanowires in the laboratory, where some Majorana splitting is inevitable because 
of the wavefunction overlap between the Majorana zero modes localized at the two ends 
of the finite wire.  The physical mechanism enabling the existence of topological visibility
is dissipation or level broadening invariably present in all real systems.  
In particular, this broadening must exceed the Majoana energy splitting for 
the system to behave 'topologically' (i.e. for the topological visibility to be negative).
But this dissipative broadening also suppresses the value of ZBCP below the 
canonically quantized value of $2e^2/h$ in the topological phase and reduces 
the magnitude of the topological visibility below unity.  For braiding experiments
of the future, a part of this dissipation arises from the finite speed of 
braiding itself which gives rise to an energy broadening, and this broadening 
must exceed the Majorana splitting energy for the system to behave as a non-Abelian 
system. Although our numerical results (when compared with tunnel conductance measurements)
indicate that some dissipative broadening must be present in the real systems, we do not 
investigate in the current work the possible physical mechanisms producing such dissipation.
At this stage, our inclusion of dissipative broadening in the theory is phenomenological, 
and future experiments will have to determine the source of such broadening in real systems.
One possibility is that the combination of disorder and magnetic 
field in the s-wave superconductor leads to subgap fermionic states at the interface. 
Such subgap fermionic states, at finite temperature and in the presence of electron phonon coupling  
can lead to the creation of a fermion bath that would have the same form assumed in this paper.
 
This work is supported by Microsoft Q, LPS-MPO-CMTC, and JQI-NSF-PFC.
We acknowledge valuable discussions with L. Kouwenhoven.

\appendix
\section{Tunneling conductance and topological
visibility from S-matrix}
Tunneling conductance is a local measurement 
at the normal lead -superconducting nanowire (see Fig.~\ref{cartoon}) junction, and one 
may calculate it theoretically by assuming both the lead and nanowire to extend 
semi-infinitely and coupled together at the so-called
Normal metal-Superconductor (NS) 
junction via a tunnel barrier.

The knowledge of the reflection matrix at the NS junction is sufficient to calculate 
the tunneling conductance.
The 
reflection matrix has the form
\begin{align}
 r = 
 \begin{pmatrix}
  r_{ee} & r_{eh} \\
  r_{he} & r_{hh}
 \end{pmatrix},
 \label{rmatrix}
\end{align}
where $r_{ee}$ and $r_{eh}$ are the normal and Andreev reflection amplitudes,
respectively. Here, the reflection matrix is expressed in the basis of electron and hole
scattering channels, which is called the particle-hole basis. Such a convenient 
decomposition in normal and Andreev reflection amplitudes is possible whenever the
lead Hamiltonian, $H_{\mathrm{lead}}$ (see Eq.~\eqref{Hlead}) is diagonal in the 
particle hole basis- i.e.
$[H_{\mathrm{lead}},\tau_z]=0$. For a single conducting channel, the 
tunneling conductance
to a superconductor in the NS junction is given by the 
Blonder-Tinkham-Klapwijk (BTK) formula
\cite{blonder1982transition} (in the units of $e^2/h$)
\begin{align}
 G = 1-|r_{ee}|^2 + |r_{eh}|^2.
 \label{conductance}
\end{align}
With N conducting modes in the lead, $r_{ee}$ and $r_{eh}$  acquire a matrix structure  
and 
the BTK formula is generalized to,
\begin{align}
 G &= N - \mathrm{Tr}(r_{ee}r_{ee}^{\dagger}-r_{eh}r_{eh}^{\dagger})\nonumber 
 \label{G}
\end{align}

For a periodic translationally invariant spinless
\textit{p}-wave superconductor described by a
Hamiltonian $H(k)$ in \textit{k}-space, Kitaev~\cite{Kit} defined the TI as
\begin{align}
Q_{\mathrm{Kitaev}} = \sgn(\mathrm{Pf}(iH(0))\mathrm{Pf}(iH(\pi))),
\end{align}
where Pf denotes Pfaffian operation on a matrix.
$Q_{\mathrm{Kitaev}}=-1$ implies that the system is in a topological phase i.e.
if the same Hamiltonian were to  describe a finite chain 
with an open boundary 
condition, the system edges will host non-Abelian Majorana zero modes.
For an open finite wire geometry, Akhmerov \textit{et. al.}~\cite{Akh} provided the
following generalization for the TI in terms of the reflection
matrix:
\begin{align}
 Q_0 = \mathrm{sgn}(\det (r)).
 \label{eq:TI}
\end{align}
It was argued in the main body of the paper that in presence of dissipation,
a more useful quantity to characterize topological properties of the system is 
TV-- a quantity closely related to scattering matrix TI~\eqref{eq:TI}, defined
as
\begin{align}
 Q = \det (r).
\end{align}

To justify this expression for the TV, 
which we use in our numerical work,
consider the particle-hole symmetry
of the superconducting Bogoliubov-de Gennes (BdG) Hamiltonian i.e.,
\begin{align}
\Pi H_{\mathrm{BdG}}\Pi^{-1} = -H_{\mathrm{BdG}},
\end{align}
where $\Pi = \tau_xC$ with $C$ being the complex conjugation
operator.
This leads to the following constraint on the reflection matrix,
\begin{align}
 \tau_x r \tau_x = r^*,
 \label{rPH}
\end{align}
which implies
\begin{align}
 \det (r) = \det (r)^* .
 \label{realDet}
\end{align}
Note that we have implicitly assumed the voltage bias, $V$, to be zero.
For finite $V$, the particle-hole constraint on the voltage-dependent reflection
matrix $r(V)$ takes the form $\tau_x r(V) \tau_x = r(-V)^*$.
When the
voltage bias is less than the superconducting gap ($eV< \Delta$),
the transmission through the nanowire is zero as there are no extended
states. Therefore the
reflection matrix $r$ is unitary i.e., $rr^{\dagger}=1$. This implies
\begin{align}
 \mathrm{Tr}(r_{ee}r_{ee}^{\dagger} + r_{eh}r_{eh}^{\dagger}) = 
 \mathrm{Tr}(r_{hh}r_{hh}^{\dagger} + r_{he}r_{he}^{\dagger}) = N
 \label{unitaryR}
\end{align}
and that the absolute value of the determinant of reflection matrix satisfies
\begin{align}
 |\det(r)| = 1.
\end{align}
Combined with the particle-hole symmetry constraint
of $r$, 
we get $\det (r)=\pm 1$. In other words we have shown that whenever
reflection matrix $r$ respects unitarity and particle-hole symmetry,
the TI (defined as $\mathrm{sgn}(\det(r))$) is equal to TV (defined as $\det(r)$), 
i.e. $Q_0 = Q$.
An ideal system with MZMs is characterized by $\det (r)
= -1$ (and non-topological trivial phase is characterized by $\det (r) = 1$)
and also is associated with quantized ZBCP at $2e^2/h$.
The only way to change the value of $\det(r)$ is to break the
unitarity by closing the topological gap. 
Note that by substituting Eq.~\eqref{unitaryR} in Eq.~\eqref{G} and using the
unitarity of the reflection matrix
one can show, 
\begin{align}
 G &= 2\mathrm{Tr}(r_{eh}r_{eh}^{\dagger}), \\
 \mathrm{Tr}(r_{eh}r_{eh}^{\dagger}) &= \mathrm{Tr}(r_{he}r_{he}^{\dagger}) \label{reh}.
\end{align}
Moreover, particle-hole symmetry of $r$ implies
\begin{align}
 r_{eh}(V) = r_{he}^{\dagger}(-V). \label{rehph}
\end{align}
Finally, using Eqs.~\eqref{rehph} and \eqref{reh} we arrive at
\begin{align}
 G(V) = G(-V).
\end{align}
So the  unitarity and particle-hole symmetry of $r$ guarantee that 
the in-gap conductance is symmetric about zero bias.
For a finite system,
any MZM would be split in energy by $\delta$
because of the inevitable MZM overlap from the two ends 
(which could be exponentially small, but never zero for a finite wire). 
Strictly at zero energy there would be no
BdG eigenstate in the nanowire 
rendering an incoming electron to be totally reflected with
$\det (r)=1$. We would infer, based on this argument, that all 
finite systems irrespective of whether they
host MZMs or not are non-topological. 
This is similar to the statement in an entirely different context 
that no finite system can have a phase transition, which is only 
a property of the infinite volume thermodynamic limit.  
In reality, other (nonuniversal) cut-offs in energy and length scales 
of the problem become important as the system size increases, and 
eventually finite and infinite systems behave in the same manner. 
For the nanowire MZM problem, this arises from the energy broadening 
inherent in any realistic system, which renders the split hybridized 
nonzero energy peaks into a broadened midgap peak with a finite weight at zero energy.  
Thus, the split resonances at sharp nonzero energies 
become a broad peak around zero energy with a finite width.
Without such a dissipative broadening process, the splitting 
of the MZMs invariably present in any real system with finite 
wire length will always lead to precisely zero conductance at 
zero energy since the MZMs are 
now always shifted from zero energy due to Majorana splitting.

We account for finite lifetime of the quasiparticle due to various inelastic scattering 
mechanisms such as phonons and magnetic moments through an onsite imaginary term in the
Hamiltonian. 
We emphasize that without this broadening, a finite wire 
can never have a true zero energy mode, 
and the system is by definition always in the trivial phase!
The resultant broadening due to the
onsite imaginary term in the Hamiltonian is given by $\Gamma $.

\section{Numerical calculation of topological visibility}
To calculate the TV, we 
numerically compute the real part of the determinant of the reflection matrix $r$ and 
discard the small imaginary part of the determinant which is a numerical artifact as can be 
seen from the fact that the particle-hole symmetry forces the determinant of the 
reflection matrix to be real which was
pointed out in Appendix A.
For all calculations involving the leads, the following 
set of parameters are chosen: $\mu_o B_{lead} = 2\mathrm{K}$ and
$\mu_{lead} = 6.9\mathrm{K}$.

Special care must be taken in calculating the TV. When the  submatrices of the
reflection matrix ($r_{ee}, r_{eh} , r_{he}$ and $r_{hh}$) are called in KWANT,
the individual submatrix outputs
do not satisfy the particle-hole constraint given by Eq.~\eqref{rPH}. The particle-hole
symmetry was restored in the following way in our calculations.
Since 
the lead parameters were chosen to have two 
incoming and two outgoing modes at zero energy, in every lead
for each participating mode $m_1,m_2$ at zero energy,
incoming and outgoing wavefunctions have a generic two component structure
\begin{align}
 \Psi = \begin{pmatrix}
                                 \psi_1 \\
                                 \psi_2
                                \end{pmatrix}. \non
\end{align}
We compute $\alpha = \mathrm{max}(\psi_1,\psi_2)$ at the end site of the lead and define the
phase of the wavefunction to be $\phi= \alpha/|\alpha|$. The arbitrary phase of the reflection
matrix is rectified by multiplying $\det (r)$ by following string of phases
\begin{align}
\phi_{in,e}^{m_1}\phi_{in,e}^{m_2}\phi_{in,h}^{m_1}\phi_{in,h}^{m_2}
\phi_{out,e}^{m_1}\phi_{out,e}^{m_2}\phi_{out,h}^{m_1}\phi_{out,h}^{m_2},
\end{align}
where $in$,$out$ stand for incoming and outgoing modes, $e$,$h$ stand for 
electron and hole and $m_1$,$m_2$ are the two modes.
These subtle numerical manipulations are essential in ensuring that the mode 
functions used by KWANT are particle-hole symmetric. The particle-hole symmetry of the basis 
is key to ensure particle-hole symmetry of the scattering
matrix that is required for the proper evaluation of the scattering matrix topological visibility.

\bibliography{ref}

\begin{thebibliography}{86}
\expandafter\ifx\csname natexlab\endcsname\relax\def\natexlab#1{#1}\fi
\expandafter\ifx\csname bibnamefont\endcsname\relax
  \def\bibnamefont#1{#1}\fi
\expandafter\ifx\csname bibfnamefont\endcsname\relax
  \def\bibfnamefont#1{#1}\fi
\expandafter\ifx\csname citenamefont\endcsname\relax
  \def\citenamefont#1{#1}\fi
\expandafter\ifx\csname url\endcsname\relax
  \def\url#1{\texttt{#1}}\fi
\expandafter\ifx\csname urlprefix\endcsname\relax\def\urlprefix{URL }\fi
\providecommand{\bibinfo}[2]{#2}
\providecommand{\eprint}[2][]{\url{#2}}

\bibitem[{\citenamefont{{V. Mourik, K. Zuo, S. M. Frolov, S. R. Plissard, E. P.
  A. M. Bakkers and L. P. Kouwenhoven.}}(2012)}]{Mou}
\bibinfo{author}{\bibnamefont{{V. Mourik, K. Zuo, S. M. Frolov, S. R. Plissard,
  E. P. A. M. Bakkers and L. P. Kouwenhoven.}}}, \bibinfo{journal}{Science}
  \textbf{\bibinfo{volume}{336}}, \bibinfo{pages}{1003} (\bibinfo{year}{2012}).

\bibitem[{\citenamefont{{Jay D. Sau, Roman M. Lutchyn, Sumanta Tewari, and S.
  Das Sarma.}}(2010)}]{Jay}
\bibinfo{author}{\bibnamefont{{Jay D. Sau, Roman M. Lutchyn, Sumanta Tewari,
  and S. Das Sarma.}}}, \bibinfo{journal}{Phys. Rev. Lett.}
  \textbf{\bibinfo{volume}{104}}, \bibinfo{pages}{040502}
  (\bibinfo{year}{2010}).

\bibitem[{\citenamefont{{Roman M. Lutchyn, Jay D. Sau, and S. Das
  Sarma.}}(2010)}]{Rom}
\bibinfo{author}{\bibnamefont{{Roman M. Lutchyn, Jay D. Sau, and S. Das
  Sarma.}}}, \bibinfo{journal}{Phys. Rev. Lett.}
  \textbf{\bibinfo{volume}{105}}, \bibinfo{pages}{077001}
  (\bibinfo{year}{2010}).

\bibitem[{\citenamefont{{Jason Alicea}}(2010)}]{Jas1}
\bibinfo{author}{\bibnamefont{{Jason Alicea}}}, \bibinfo{journal}{Phys. Rev. B}
  \textbf{\bibinfo{volume}{81}}, \bibinfo{pages}{125318}
  (\bibinfo{year}{2010}).

\bibitem[{\citenamefont{{Jay D. Sau, Sumanta Tewari, Roman M. Lutchyn, Tudor D.
  Stanescu, and S. Das Sarma.}}(2010)}]{Jay1}
\bibinfo{author}{\bibnamefont{{Jay D. Sau, Sumanta Tewari, Roman M. Lutchyn,
  Tudor D. Stanescu, and S. Das Sarma.}}}, \bibinfo{journal}{Phys. Rev. B}
  \textbf{\bibinfo{volume}{82}}, \bibinfo{pages}{214509}
  (\bibinfo{year}{2010}).

\bibitem[{\citenamefont{{Yuval Oreg, Gil Refael and Felix von
  Oppen.}}(2010)}]{Yuv}
\bibinfo{author}{\bibnamefont{{Yuval Oreg, Gil Refael and Felix von Oppen.}}},
  \bibinfo{journal}{Phys. Rev. Lett.} \textbf{\bibinfo{volume}{105}},
  \bibinfo{pages}{177002} (\bibinfo{year}{2010}).

\bibitem[{\citenamefont{Liu et~al.}(2015)\citenamefont{Liu, Sau, and
  Das~Sarma}}]{liu2015universal}
\bibinfo{author}{\bibfnamefont{X.}~\bibnamefont{Liu}},
  \bibinfo{author}{\bibfnamefont{J.~D.} \bibnamefont{Sau}}, \bibnamefont{and}
  \bibinfo{author}{\bibfnamefont{S.}~\bibnamefont{Das~Sarma}},
  \bibinfo{journal}{Physical Review B} \textbf{\bibinfo{volume}{92}},
  \bibinfo{pages}{014513} (\bibinfo{year}{2015}).

\bibitem[{\citenamefont{{M. T. Deng, C. L. Yu, G. Y. Huang, M. Larsson, P.
  Caroff, and H. Q. Xu.}}(2012)}]{Den}
\bibinfo{author}{\bibnamefont{{M. T. Deng, C. L. Yu, G. Y. Huang, M. Larsson,
  P. Caroff, and H. Q. Xu.}}}, \bibinfo{journal}{Nano Lett.}
  \textbf{\bibinfo{volume}{12}}, \bibinfo{pages}{6414} (\bibinfo{year}{2012}).

\bibitem[{\citenamefont{{Leonid P Rokhinson, Xinyu Liu, and Jacek K
  Furdyna.}}(2012)}]{Leo}
\bibinfo{author}{\bibnamefont{{Leonid P Rokhinson, Xinyu Liu, and Jacek K
  Furdyna.}}}, \bibinfo{journal}{Nat. Phys.} \textbf{\bibinfo{volume}{8}},
  \bibinfo{pages}{795} (\bibinfo{year}{2012}).

\bibitem[{\citenamefont{{Anindya Das, Yuval Ronen, Yonatan Most, Yuval Oreg,
  Moty Heiblum and Hadas Shtrikman.}}(2012)}]{Das}
\bibinfo{author}{\bibnamefont{{Anindya Das, Yuval Ronen, Yonatan Most, Yuval
  Oreg, Moty Heiblum and Hadas Shtrikman.}}}, \bibinfo{journal}{Nat. Phys.}
  \textbf{\bibinfo{volume}{8}}, \bibinfo{pages}{887} (\bibinfo{year}{2012}).

\bibitem[{\citenamefont{{A. D. K. Finck, D. J. Van Harlingen, P. K. Mohseni, K.
  Jung, and X. Li}}(2013)}]{Fin}
\bibinfo{author}{\bibnamefont{{A. D. K. Finck, D. J. Van Harlingen, P. K.
  Mohseni, K. Jung, and X. Li}}}, \bibinfo{journal}{Phys. Rev. Lett}
  \textbf{\bibinfo{volume}{110}}, \bibinfo{pages}{126406}
  (\bibinfo{year}{2013}).

\bibitem[{\citenamefont{{H. O. H. Churchill, V. Fatemi, K. Grove-Rasmussen, M.
  T. Deng, P. Caroff, H. Q. Xu, and C. M. Marcus.}}(2013)}]{Chu}
\bibinfo{author}{\bibnamefont{{H. O. H. Churchill, V. Fatemi, K.
  Grove-Rasmussen, M. T. Deng, P. Caroff, H. Q. Xu, and C. M. Marcus.}}},
  \bibinfo{journal}{Phys. Rev. B} \textbf{\bibinfo{volume}{87}},
  \bibinfo{pages}{241401} (\bibinfo{year}{2013}).

\bibitem[{\citenamefont{{W. Chang, S. M. Albrecht, T. S. Jespersen, F.
  Kuemmeth, P. Krogstrup, J. Nygård and C. M. Marcus}}(2015)}]{Cha}
\bibinfo{author}{\bibnamefont{{W. Chang, S. M. Albrecht, T. S. Jespersen, F.
  Kuemmeth, P. Krogstrup, J. Nygård and C. M. Marcus}}},
  \bibinfo{journal}{Nature Nanotechnology} \textbf{\bibinfo{volume}{10}},
  \bibinfo{pages}{232} (\bibinfo{year}{2015}).

\bibitem[{\citenamefont{{Alexei Kitaev}}(2001)}]{Kit}
\bibinfo{author}{\bibnamefont{{Alexei Kitaev}}}, \bibinfo{journal}{Phys. Usp.}
  \textbf{\bibinfo{volume}{44}}, \bibinfo{pages}{131} (\bibinfo{year}{2001}).

\bibitem[{\citenamefont{{K. Sengupta, I. Zutic, H.-J. Kwon, V. M. Yakovenko,
  and S. Das Sarma}}(2001)}]{Sen}
\bibinfo{author}{\bibnamefont{{K. Sengupta, I. Zutic, H.-J. Kwon, V. M.
  Yakovenko, and S. Das Sarma}}}, \bibinfo{journal}{Phys. Rev. B}
  \textbf{\bibinfo{volume}{63}}, \bibinfo{pages}{144531}
  (\bibinfo{year}{2001}).

\bibitem[{\citenamefont{Law et~al.}(2009)\citenamefont{Law, Lee, and
  Ng}}]{law2009majorana}
\bibinfo{author}{\bibfnamefont{K.~T.} \bibnamefont{Law}},
  \bibinfo{author}{\bibfnamefont{P.~A.} \bibnamefont{Lee}}, \bibnamefont{and}
  \bibinfo{author}{\bibfnamefont{T.~K.} \bibnamefont{Ng}},
  \bibinfo{journal}{Phys. Rev. Lett.} \textbf{\bibinfo{volume}{103}},
  \bibinfo{pages}{237001} (\bibinfo{year}{2009}).

\bibitem[{\citenamefont{{Chetan Nayak, Steven H. Simon, Ady Stern, Michael
  Freedman, and S. Das Sarma.}}(2008)}]{Che}
\bibinfo{author}{\bibnamefont{{Chetan Nayak, Steven H. Simon, Ady Stern,
  Michael Freedman, and S. Das Sarma.}}}, \bibinfo{journal}{Rev. Mod. Phys.}
  \textbf{\bibinfo{volume}{80}}, \bibinfo{pages}{1083} (\bibinfo{year}{2008}).

\bibitem[{\citenamefont{{Jason Alicea}}(2012)}]{Jas}
\bibinfo{author}{\bibnamefont{{Jason Alicea}}}, \bibinfo{journal}{Rep. Prog.
  Phys.} \textbf{\bibinfo{volume}{75}}, \bibinfo{pages}{076501}
  (\bibinfo{year}{2012}).

\bibitem[{\citenamefont{{C. W. J. Beenakker}}(2013)}]{Bee1}
\bibinfo{author}{\bibnamefont{{C. W. J. Beenakker}}}, \bibinfo{journal}{Annu.
  Rev. Con. Mat. Phys.} \textbf{\bibinfo{volume}{4}}, \bibinfo{pages}{113}
  (\bibinfo{year}{2013}).

\bibitem[{\citenamefont{{Martin Leijnse and Karsten Flensberg.}}(2012)}]{Kar}
\bibinfo{author}{\bibnamefont{{Martin Leijnse and Karsten Flensberg.}}},
  \bibinfo{journal}{Semicond. Sci. Technol.} \textbf{\bibinfo{volume}{27}},
  \bibinfo{pages}{124003} (\bibinfo{year}{2012}).

\bibitem[{\citenamefont{{Steven R. Elliott and Marcel Franz}}(2015)}]{Mar}
\bibinfo{author}{\bibnamefont{{Steven R. Elliott and Marcel Franz}}},
  \bibinfo{journal}{Rev. Mod. Phys.} \textbf{\bibinfo{volume}{87}},
  \bibinfo{pages}{137} (\bibinfo{year}{2015}).

\bibitem[{\citenamefont{{T D Stanescu and S Tewari.}}(2013)}]{Sta}
\bibinfo{author}{\bibnamefont{{T D Stanescu and S Tewari.}}},
  \bibinfo{journal}{J. Phys. Condens. Matter} \textbf{\bibinfo{volume}{25}},
  \bibinfo{pages}{233201} (\bibinfo{year}{2013}).

\bibitem[{\citenamefont{{S. Das Sarma, M. Freedman, and C. Nayak}}(2006)}]{San}
\bibinfo{author}{\bibnamefont{{S. Das Sarma, M. Freedman, and C. Nayak}}},
  \bibinfo{journal}{Physics Today} \textbf{\bibinfo{volume}{59}},
  \bibinfo{pages}{32} (\bibinfo{year}{2006}).

\bibitem[{\citenamefont{Das~Sarma
  et~al.}(2015{\natexlab{a}})\citenamefont{Das~Sarma, Freedman, and
  Nayak}}]{sarma2015majorana}
\bibinfo{author}{\bibfnamefont{S.}~\bibnamefont{Das~Sarma}},
  \bibinfo{author}{\bibfnamefont{M.}~\bibnamefont{Freedman}}, \bibnamefont{and}
  \bibinfo{author}{\bibfnamefont{C.}~\bibnamefont{Nayak}},
  \bibinfo{journal}{Npj Quantum Information} \textbf{\bibinfo{volume}{1}},
  \bibinfo{pages}{15001} (\bibinfo{year}{2015}{\natexlab{a}}).

\bibitem[{\citenamefont{{Dmitry Bagrets and Alexander Altland}}(2012)}]{Alt1}
\bibinfo{author}{\bibnamefont{{Dmitry Bagrets and Alexander Altland}}},
  \bibinfo{journal}{Phys. Rev. Lett.} \textbf{\bibinfo{volume}{109}},
  \bibinfo{pages}{227005} (\bibinfo{year}{2012}).

\bibitem[{\citenamefont{{Patrick Neven, Dmitry Bagrets, and Alexander
  Altland.}}(2013)}]{Pat}
\bibinfo{author}{\bibnamefont{{Patrick Neven, Dmitry Bagrets, and Alexander
  Altland.}}}, \bibinfo{journal}{New J. Phys.} \textbf{\bibinfo{volume}{15}},
  \bibinfo{pages}{055019} (\bibinfo{year}{2013}).

\bibitem[{\citenamefont{{Eduardo J. H. Lee, Xiaocheng Jiang, Manuel Houzet,
  Ramon Aguado, Charles M. Lieber, and Silvano De Franceschi}}(2013)}]{Edu}
\bibinfo{author}{\bibnamefont{{Eduardo J. H. Lee, Xiaocheng Jiang, Manuel
  Houzet, Ramon Aguado, Charles M. Lieber, and Silvano De Franceschi}}},
  \bibinfo{journal}{Nat Nano} \textbf{\bibinfo{volume}{9}}, \bibinfo{pages}{79}
  (\bibinfo{year}{2013}).

\bibitem[{\citenamefont{{Falko Pientka, Graham Kells, Alessandro Romito, Piet
  W. Brouwer and Felix von Oppen.}}(2012)}]{Fal}
\bibinfo{author}{\bibnamefont{{Falko Pientka, Graham Kells, Alessandro Romito,
  Piet W. Brouwer and Felix von Oppen.}}}, \bibinfo{journal}{Phys. Rev. Lett.}
  \textbf{\bibinfo{volume}{109}}, \bibinfo{pages}{227006}
  (\bibinfo{year}{2012}).

\bibitem[{\citenamefont{{G. Kells, D. Meidan, and P. W. Brouwer}}(2012)}]{Kel}
\bibinfo{author}{\bibnamefont{{G. Kells, D. Meidan, and P. W. Brouwer}}},
  \bibinfo{journal}{Phys. Rev. B} \textbf{\bibinfo{volume}{85}},
  \bibinfo{pages}{060507} (\bibinfo{year}{2012}).

\bibitem[{\citenamefont{{Jie Liu, Andrew C. Potter, K. T. Law, and Patrick A.
  Lee.}}(2012)}]{Jie}
\bibinfo{author}{\bibnamefont{{Jie Liu, Andrew C. Potter, K. T. Law, and
  Patrick A. Lee.}}}, \bibinfo{journal}{Phys. Rev. Lett.}
  \textbf{\bibinfo{volume}{109}}, \bibinfo{pages}{267002}
  (\bibinfo{year}{2012}).

\bibitem[{\citenamefont{Pikulin et~al.}(2012)\citenamefont{Pikulin, Dahlhaus,
  Wimmer, Schomerus, and Beenakker}}]{pikulin2012weakanti}
\bibinfo{author}{\bibfnamefont{D.~I.} \bibnamefont{Pikulin}},
  \bibinfo{author}{\bibfnamefont{J.~P.} \bibnamefont{Dahlhaus}},
  \bibinfo{author}{\bibfnamefont{M.}~\bibnamefont{Wimmer}},
  \bibinfo{author}{\bibfnamefont{H.}~\bibnamefont{Schomerus}},
  \bibnamefont{and} \bibinfo{author}{\bibfnamefont{C.~W.~J.}
  \bibnamefont{Beenakker}}, \bibinfo{journal}{New Journal of Physics}
  \textbf{\bibinfo{volume}{14}}, \bibinfo{pages}{125011}
  (\bibinfo{year}{2012}).

\bibitem[{\citenamefont{Flensberg}(2010)}]{flensberg2010tunneling}
\bibinfo{author}{\bibfnamefont{K.}~\bibnamefont{Flensberg}},
  \bibinfo{journal}{Physical Review B} \textbf{\bibinfo{volume}{82}},
  \bibinfo{pages}{180516} (\bibinfo{year}{2010}).

\bibitem[{\citenamefont{Wimmer et~al.}(2011)\citenamefont{Wimmer, Akhmerov,
  Dahlhaus, and Beenakker}}]{wimmer2011quantum}
\bibinfo{author}{\bibfnamefont{M.}~\bibnamefont{Wimmer}},
  \bibinfo{author}{\bibfnamefont{A.~R.} \bibnamefont{Akhmerov}},
  \bibinfo{author}{\bibfnamefont{J.~P.} \bibnamefont{Dahlhaus}},
  \bibnamefont{and} \bibinfo{author}{\bibfnamefont{C.~W.~J.}
  \bibnamefont{Beenakker}}, \bibinfo{journal}{New Journal of Physics}
  \textbf{\bibinfo{volume}{13}}, \bibinfo{pages}{053016}
  (\bibinfo{year}{2011}).

\bibitem[{\citenamefont{Lin et~al.}(2012)\citenamefont{Lin, Sau, and
  Das~Sarma}}]{lin2012zero}
\bibinfo{author}{\bibfnamefont{C.-H.} \bibnamefont{Lin}},
  \bibinfo{author}{\bibfnamefont{J.~D.} \bibnamefont{Sau}}, \bibnamefont{and}
  \bibinfo{author}{\bibfnamefont{S.}~\bibnamefont{Das~Sarma}},
  \bibinfo{journal}{Phys. Rev. B} \textbf{\bibinfo{volume}{86}},
  \bibinfo{pages}{224511} (\bibinfo{year}{2012}).

\bibitem[{\citenamefont{Zhang et~al.}(2016)\citenamefont{Zhang, G{\"u}l,
  Conesa-Boj, Zuo, Mourik, de~Vries, van Veen, van Woerkom, Nowak, Wimmer
  et~al.}}]{delftballistic}
\bibinfo{author}{\bibfnamefont{H.}~\bibnamefont{Zhang}},
  \bibinfo{author}{\bibfnamefont{{\"O}.}~\bibnamefont{G{\"u}l}},
  \bibinfo{author}{\bibfnamefont{S.}~\bibnamefont{Conesa-Boj}},
  \bibinfo{author}{\bibfnamefont{K.}~\bibnamefont{Zuo}},
  \bibinfo{author}{\bibfnamefont{V.}~\bibnamefont{Mourik}},
  \bibinfo{author}{\bibfnamefont{F.~K.} \bibnamefont{de~Vries}},
  \bibinfo{author}{\bibfnamefont{J.}~\bibnamefont{van Veen}},
  \bibinfo{author}{\bibfnamefont{D.~J.} \bibnamefont{van Woerkom}},
  \bibinfo{author}{\bibfnamefont{M.~P.} \bibnamefont{Nowak}},
  \bibinfo{author}{\bibfnamefont{M.}~\bibnamefont{Wimmer}},
  \bibnamefont{et~al.}, \bibinfo{journal}{arXiv preprint arXiv:1603.04069}
  (\bibinfo{year}{2016}).

\bibitem[{\citenamefont{{Meng Cheng, Roman M. Lutchyn, Victor Galitski and S.
  Das Sarma}}(2009)}]{Meng}
\bibinfo{author}{\bibnamefont{{Meng Cheng, Roman M. Lutchyn, Victor Galitski
  and S. Das Sarma}}}, \bibinfo{journal}{Phys. Rev. Lett.}
  \textbf{\bibinfo{volume}{103}}, \bibinfo{pages}{107001}
  (\bibinfo{year}{2009}).

\bibitem[{\citenamefont{{Meng Cheng, Roman M. Lutchyn, Victor Galitski, S. Das
  Sarma}}(2010)}]{Meng1}
\bibinfo{author}{\bibnamefont{{Meng Cheng, Roman M. Lutchyn, Victor Galitski,
  S. Das Sarma}}}, \bibinfo{journal}{Phys. Rev. B}
  \textbf{\bibinfo{volume}{82}}, \bibinfo{pages}{094504}
  (\bibinfo{year}{2010}).

\bibitem[{\citenamefont{{Diego Rainis, Luka Trifunovic, Jelena Klinovaja and
  Daniel Loss}}(2013)}]{Rai}
\bibinfo{author}{\bibnamefont{{Diego Rainis, Luka Trifunovic, Jelena Klinovaja
  and Daniel Loss}}}, \bibinfo{journal}{Phys. Rev. B}
  \textbf{\bibinfo{volume}{87}}, \bibinfo{pages}{024515}
  (\bibinfo{year}{2013}).

\bibitem[{\citenamefont{{S. Das Sarma, Jay D Sau and Tudor D
  Stanescu}}(2012)}]{Das1}
\bibinfo{author}{\bibnamefont{{S. Das Sarma, Jay D Sau and Tudor D Stanescu}}},
  \bibinfo{journal}{Phys. Rev. B} \textbf{\bibinfo{volume}{86}},
  \bibinfo{pages}{220506} (\bibinfo{year}{2012}).

\bibitem[{\citenamefont{Takei et~al.}(2013)\citenamefont{Takei, Fregoso, Hui,
  Lobos, and Das~Sarma}}]{takei2013soft}
\bibinfo{author}{\bibfnamefont{S.}~\bibnamefont{Takei}},
  \bibinfo{author}{\bibfnamefont{B.~M.} \bibnamefont{Fregoso}},
  \bibinfo{author}{\bibfnamefont{H.-Y.} \bibnamefont{Hui}},
  \bibinfo{author}{\bibfnamefont{A.~M.} \bibnamefont{Lobos}}, \bibnamefont{and}
  \bibinfo{author}{\bibfnamefont{S.}~\bibnamefont{Das~Sarma}},
  \bibinfo{journal}{Phys. Rev. Lett.} \textbf{\bibinfo{volume}{110}},
  \bibinfo{pages}{186803} (\bibinfo{year}{2013}).

\bibitem[{\citenamefont{{Jay D. Sau, S. Das Sarma}}(2013)}]{SauSarma}
\bibinfo{author}{\bibnamefont{{Jay D. Sau, S. Das Sarma}}},
  \bibinfo{journal}{Phys. Rev. B.} \textbf{\bibinfo{volume}{88}},
  \bibinfo{pages}{064506} (\bibinfo{year}{2013}).

\bibitem[{\citenamefont{Brouwer et~al.}(2011)\citenamefont{Brouwer, Duckheim,
  Romito, and von Oppen}}]{brouwer2011topological}
\bibinfo{author}{\bibfnamefont{P.~W.} \bibnamefont{Brouwer}},
  \bibinfo{author}{\bibfnamefont{M.}~\bibnamefont{Duckheim}},
  \bibinfo{author}{\bibfnamefont{A.}~\bibnamefont{Romito}}, \bibnamefont{and}
  \bibinfo{author}{\bibfnamefont{F.}~\bibnamefont{von Oppen}},
  \bibinfo{journal}{Phys. Rev. B} \textbf{\bibinfo{volume}{84}},
  \bibinfo{pages}{144526} (\bibinfo{year}{2011}).

\bibitem[{\citenamefont{{Piet W. Brouwer, Mathias Duckheim, Alessandro Romito,
  and Felix von Oppen}}(2011)}]{Pie}
\bibinfo{author}{\bibnamefont{{Piet W. Brouwer, Mathias Duckheim, Alessandro
  Romito, and Felix von Oppen}}}, \bibinfo{journal}{Phys. Rev. Lett.}
  \textbf{\bibinfo{volume}{107}}, \bibinfo{pages}{196804}
  (\bibinfo{year}{2011}).

\bibitem[{\citenamefont{Prada et~al.}(2012)\citenamefont{Prada, San-Jose, and
  Aguado}}]{prada2012transport}
\bibinfo{author}{\bibfnamefont{E.}~\bibnamefont{Prada}},
  \bibinfo{author}{\bibfnamefont{P.}~\bibnamefont{San-Jose}}, \bibnamefont{and}
  \bibinfo{author}{\bibfnamefont{R.}~\bibnamefont{Aguado}},
  \bibinfo{journal}{Phys. Rev. B} \textbf{\bibinfo{volume}{86}},
  \bibinfo{pages}{180503} (\bibinfo{year}{2012}).

\bibitem[{\citenamefont{Chevallier et~al.}(2013)\citenamefont{Chevallier,
  Simon, and Bena}}]{chevallier2013andreev}
\bibinfo{author}{\bibfnamefont{D.}~\bibnamefont{Chevallier}},
  \bibinfo{author}{\bibfnamefont{P.}~\bibnamefont{Simon}}, \bibnamefont{and}
  \bibinfo{author}{\bibfnamefont{C.}~\bibnamefont{Bena}},
  \bibinfo{journal}{Phys. Rev. B} \textbf{\bibinfo{volume}{88}},
  \bibinfo{pages}{165401} (\bibinfo{year}{2013}).

\bibitem[{\citenamefont{{C. Marcus}}()}]{MarUnp}
\bibinfo{author}{\bibnamefont{{C. Marcus}}}, \bibinfo{note}{unpublished}.

\bibitem[{\citenamefont{Lobos and Das~Sarma}(2015)}]{lobos2015tunneling}
\bibinfo{author}{\bibfnamefont{A.~M.} \bibnamefont{Lobos}} \bibnamefont{and}
  \bibinfo{author}{\bibfnamefont{S.}~\bibnamefont{Das~Sarma}},
  \bibinfo{journal}{New Journal of Physics} \textbf{\bibinfo{volume}{17}},
  \bibinfo{pages}{065010} (\bibinfo{year}{2015}).

\bibitem[{\citenamefont{Fregoso et~al.}(2013)\citenamefont{Fregoso, Lobos, and
  Das~Sarma}}]{fregoso2013electrical}
\bibinfo{author}{\bibfnamefont{B.~M.} \bibnamefont{Fregoso}},
  \bibinfo{author}{\bibfnamefont{A.~M.} \bibnamefont{Lobos}}, \bibnamefont{and}
  \bibinfo{author}{\bibfnamefont{S.}~\bibnamefont{Das~Sarma}},
  \bibinfo{journal}{Physical Review B} \textbf{\bibinfo{volume}{88}},
  \bibinfo{pages}{180507} (\bibinfo{year}{2013}).

\bibitem[{\citenamefont{{N. Read and D. Green.}}(2000)}]{Rea}
\bibinfo{author}{\bibnamefont{{N. Read and D. Green.}}},
  \bibinfo{journal}{Phys. Rev. B} \textbf{\bibinfo{volume}{61}},
  \bibinfo{pages}{10267} (\bibinfo{year}{2000}).

\bibitem[{\citenamefont{{D.A. Ivanov}}(2001)}]{Iva}
\bibinfo{author}{\bibnamefont{{D.A. Ivanov}}}, \bibinfo{journal}{Phys. Rev.
  Lett} \textbf{\bibinfo{volume}{86}}, \bibinfo{pages}{268}
  (\bibinfo{year}{2001}).

\bibitem[{\citenamefont{{Jason Alicea, Yuval Oreg, Gil Refael, Felix von Oppen
  and Matthew P. A. Fisher}}(2011)}]{Jas2}
\bibinfo{author}{\bibnamefont{{Jason Alicea, Yuval Oreg, Gil Refael, Felix von
  Oppen and Matthew P. A. Fisher}}}, \bibinfo{journal}{Nat. Phys.}
  \textbf{\bibinfo{volume}{7}}, \bibinfo{pages}{412} (\bibinfo{year}{2011}).

\bibitem[{\citenamefont{Sau et~al.}(2011{\natexlab{a}})\citenamefont{Sau,
  Tewari, and Das~Sarma}}]{sau2011QComp}
\bibinfo{author}{\bibfnamefont{J.~D.} \bibnamefont{Sau}},
  \bibinfo{author}{\bibfnamefont{S.}~\bibnamefont{Tewari}}, \bibnamefont{and}
  \bibinfo{author}{\bibfnamefont{S.}~\bibnamefont{Das~Sarma}},
  \bibinfo{journal}{Phys. Rev. B} \textbf{\bibinfo{volume}{84}},
  \bibinfo{pages}{085109} (\bibinfo{year}{2011}{\natexlab{a}}).

\bibitem[{\citenamefont{Clarke et~al.}(2011)\citenamefont{Clarke, Sau, and
  Tewari}}]{clarke2011QComp}
\bibinfo{author}{\bibfnamefont{D.~J.} \bibnamefont{Clarke}},
  \bibinfo{author}{\bibfnamefont{J.~D.} \bibnamefont{Sau}}, \bibnamefont{and}
  \bibinfo{author}{\bibfnamefont{S.}~\bibnamefont{Tewari}},
  \bibinfo{journal}{Phys. Rev. B} \textbf{\bibinfo{volume}{84}},
  \bibinfo{pages}{035120} (\bibinfo{year}{2011}).

\bibitem[{\citenamefont{Hyart et~al.}(2013)\citenamefont{Hyart, van Heck,
  Fulga, Burrello, Akhmerov, and Beenakker}}]{hyart2013flux}
\bibinfo{author}{\bibfnamefont{T.}~\bibnamefont{Hyart}},
  \bibinfo{author}{\bibfnamefont{B.}~\bibnamefont{van Heck}},
  \bibinfo{author}{\bibfnamefont{I.~C.} \bibnamefont{Fulga}},
  \bibinfo{author}{\bibfnamefont{M.}~\bibnamefont{Burrello}},
  \bibinfo{author}{\bibfnamefont{A.~R.} \bibnamefont{Akhmerov}},
  \bibnamefont{and} \bibinfo{author}{\bibfnamefont{C.~W.~J.}
  \bibnamefont{Beenakker}}, \bibinfo{journal}{Phys. Rev. B}
  \textbf{\bibinfo{volume}{88}}, \bibinfo{pages}{035121}
  (\bibinfo{year}{2013}).

\bibitem[{\citenamefont{Amorim et~al.}(2015)\citenamefont{Amorim, Ebihara,
  Yamakage, Tanaka, and Sato}}]{amorim2015majorana}
\bibinfo{author}{\bibfnamefont{C.~S.} \bibnamefont{Amorim}},
  \bibinfo{author}{\bibfnamefont{K.}~\bibnamefont{Ebihara}},
  \bibinfo{author}{\bibfnamefont{A.}~\bibnamefont{Yamakage}},
  \bibinfo{author}{\bibfnamefont{Y.}~\bibnamefont{Tanaka}}, \bibnamefont{and}
  \bibinfo{author}{\bibfnamefont{M.}~\bibnamefont{Sato}},
  \bibinfo{journal}{Phys. Rev. B} \textbf{\bibinfo{volume}{91}},
  \bibinfo{pages}{174305} (\bibinfo{year}{2015}).

\bibitem[{\citenamefont{Van~Heck et~al.}(2012)\citenamefont{Van~Heck, Akhmerov,
  Hassler, Burrello, and Beenakker}}]{van2012coulomb}
\bibinfo{author}{\bibfnamefont{B.}~\bibnamefont{Van~Heck}},
  \bibinfo{author}{\bibfnamefont{A.~R.} \bibnamefont{Akhmerov}},
  \bibinfo{author}{\bibfnamefont{F.}~\bibnamefont{Hassler}},
  \bibinfo{author}{\bibfnamefont{M.}~\bibnamefont{Burrello}}, \bibnamefont{and}
  \bibinfo{author}{\bibfnamefont{C.~W.~J.} \bibnamefont{Beenakker}},
  \bibinfo{journal}{New Journal of Physics} \textbf{\bibinfo{volume}{14}},
  \bibinfo{pages}{035019} (\bibinfo{year}{2012}).

\bibitem[{\citenamefont{Liang et~al.}(2012)\citenamefont{Liang, Wang, and
  Hu}}]{liang2012manipulation}
\bibinfo{author}{\bibfnamefont{Q.-F.} \bibnamefont{Liang}},
  \bibinfo{author}{\bibfnamefont{Z.}~\bibnamefont{Wang}}, \bibnamefont{and}
  \bibinfo{author}{\bibfnamefont{X.}~\bibnamefont{Hu}}, \bibinfo{journal}{EPL
  (Europhysics Letters)} \textbf{\bibinfo{volume}{99}}, \bibinfo{pages}{50004}
  (\bibinfo{year}{2012}).

\bibitem[{\citenamefont{Romito et~al.}(2012)\citenamefont{Romito, Alicea,
  Refael, and von Oppen}}]{romito2012manipulating}
\bibinfo{author}{\bibfnamefont{A.}~\bibnamefont{Romito}},
  \bibinfo{author}{\bibfnamefont{J.}~\bibnamefont{Alicea}},
  \bibinfo{author}{\bibfnamefont{G.}~\bibnamefont{Refael}}, \bibnamefont{and}
  \bibinfo{author}{\bibfnamefont{F.}~\bibnamefont{von Oppen}},
  \bibinfo{journal}{Physical Review B} \textbf{\bibinfo{volume}{85}},
  \bibinfo{pages}{020502} (\bibinfo{year}{2012}).

\bibitem[{\citenamefont{Halperin et~al.}(2012)\citenamefont{Halperin, Oreg,
  Stern, Refael, Alicea, and von Oppen}}]{halperin2012adiabatic}
\bibinfo{author}{\bibfnamefont{B.~I.} \bibnamefont{Halperin}},
  \bibinfo{author}{\bibfnamefont{Y.}~\bibnamefont{Oreg}},
  \bibinfo{author}{\bibfnamefont{A.}~\bibnamefont{Stern}},
  \bibinfo{author}{\bibfnamefont{G.}~\bibnamefont{Refael}},
  \bibinfo{author}{\bibfnamefont{J.}~\bibnamefont{Alicea}}, \bibnamefont{and}
  \bibinfo{author}{\bibfnamefont{F.}~\bibnamefont{von Oppen}},
  \bibinfo{journal}{Physical Review B} \textbf{\bibinfo{volume}{85}},
  \bibinfo{pages}{144501} (\bibinfo{year}{2012}).

\bibitem[{\citenamefont{Kotetes et~al.}(2013)\citenamefont{Kotetes, Sch{\"o}n,
  and Shnirman}}]{kotetes2013engineering}
\bibinfo{author}{\bibfnamefont{P.}~\bibnamefont{Kotetes}},
  \bibinfo{author}{\bibfnamefont{G.}~\bibnamefont{Sch{\"o}n}},
  \bibnamefont{and} \bibinfo{author}{\bibfnamefont{A.}~\bibnamefont{Shnirman}},
  \bibinfo{journal}{Journal of the Korean Physical Society}
  \textbf{\bibinfo{volume}{62}}, \bibinfo{pages}{1558} (\bibinfo{year}{2013}).

\bibitem[{\citenamefont{Clarke et~al.}(2015)\citenamefont{Clarke, Sau, and
  Das~Sarma}}]{clarke2015bell}
\bibinfo{author}{\bibfnamefont{D.~J.} \bibnamefont{Clarke}},
  \bibinfo{author}{\bibfnamefont{J.~D.} \bibnamefont{Sau}}, \bibnamefont{and}
  \bibinfo{author}{\bibfnamefont{S.}~\bibnamefont{Das~Sarma}},
  \bibinfo{journal}{arXiv preprint arXiv:1510.00007}  (\bibinfo{year}{2015}).

\bibitem[{\citenamefont{Liu and Lobos}(2013)}]{liu2013manipulating}
\bibinfo{author}{\bibfnamefont{X.-J.} \bibnamefont{Liu}} \bibnamefont{and}
  \bibinfo{author}{\bibfnamefont{A.~M.} \bibnamefont{Lobos}},
  \bibinfo{journal}{Phys. Rev. B} \textbf{\bibinfo{volume}{87}},
  \bibinfo{pages}{060504} (\bibinfo{year}{2013}).

\bibitem[{\citenamefont{Chiu et~al.}(2015)\citenamefont{Chiu, Vazifeh, and
  Franz}}]{chiu2015majorana}
\bibinfo{author}{\bibfnamefont{C.-K.} \bibnamefont{Chiu}},
  \bibinfo{author}{\bibfnamefont{M.}~\bibnamefont{Vazifeh}}, \bibnamefont{and}
  \bibinfo{author}{\bibfnamefont{M.}~\bibnamefont{Franz}},
  \bibinfo{journal}{EPL (Europhysics Letters)} \textbf{\bibinfo{volume}{110}},
  \bibinfo{pages}{10001} (\bibinfo{year}{2015}).

\bibitem[{\citenamefont{Sau et~al.}(2011{\natexlab{b}})\citenamefont{Sau,
  Clarke, and Tewari}}]{sau2011controlling}
\bibinfo{author}{\bibfnamefont{J.~D.} \bibnamefont{Sau}},
  \bibinfo{author}{\bibfnamefont{D.~J.} \bibnamefont{Clarke}},
  \bibnamefont{and} \bibinfo{author}{\bibfnamefont{S.}~\bibnamefont{Tewari}},
  \bibinfo{journal}{Physical Review B} \textbf{\bibinfo{volume}{84}},
  \bibinfo{pages}{094505} (\bibinfo{year}{2011}{\natexlab{b}}).

\bibitem[{\citenamefont{Sau et~al.}(2010)\citenamefont{Sau, Tewari, and
  Das~Sarma}}]{sau2010universal}
\bibinfo{author}{\bibfnamefont{J.~D.} \bibnamefont{Sau}},
  \bibinfo{author}{\bibfnamefont{S.}~\bibnamefont{Tewari}}, \bibnamefont{and}
  \bibinfo{author}{\bibfnamefont{S.}~\bibnamefont{Das~Sarma}},
  \bibinfo{journal}{Physical Review A} \textbf{\bibinfo{volume}{82}},
  \bibinfo{pages}{052322} (\bibinfo{year}{2010}).

\bibitem[{\citenamefont{{A. R. Akhmerov, J. P. Dahlhaus, F. Hassler, M. Wimmer,
  and C. W. J. Beenakker}}(2011)}]{Akh}
\bibinfo{author}{\bibnamefont{{A. R. Akhmerov, J. P. Dahlhaus, F. Hassler, M.
  Wimmer, and C. W. J. Beenakker}}}, \bibinfo{journal}{Phys. Rev. Lett}
  \textbf{\bibinfo{volume}{106}}, \bibinfo{pages}{057001}
  (\bibinfo{year}{2011}).

\bibitem[{\citenamefont{Adagideli et~al.}(2014)\citenamefont{Adagideli, Wimmer,
  and Teker}}]{adagideli2014effects}
\bibinfo{author}{\bibfnamefont{{\.I}.}~\bibnamefont{Adagideli}},
  \bibinfo{author}{\bibfnamefont{M.}~\bibnamefont{Wimmer}}, \bibnamefont{and}
  \bibinfo{author}{\bibfnamefont{A.}~\bibnamefont{Teker}},
  \bibinfo{journal}{Phys. Rev. B} \textbf{\bibinfo{volume}{89}},
  \bibinfo{pages}{144506} (\bibinfo{year}{2014}).

\bibitem[{\citenamefont{{I. C. Fulga, F. Hassler, A. R. Akhmerov and C. W. J.
  Beenakker }}(2011)}]{Ful}
\bibinfo{author}{\bibnamefont{{I. C. Fulga, F. Hassler, A. R. Akhmerov and C.
  W. J. Beenakker }}}, \bibinfo{journal}{Phys. Rev. B}
  \textbf{\bibinfo{volume}{83}}, \bibinfo{pages}{155429}
  (\bibinfo{year}{2011}).

\bibitem[{\citenamefont{Pikulin and Nazarov}(2012)}]{pikulin2012topological}
\bibinfo{author}{\bibfnamefont{D.~I.} \bibnamefont{Pikulin}} \bibnamefont{and}
  \bibinfo{author}{\bibfnamefont{Y.~V.} \bibnamefont{Nazarov}},
  \bibinfo{journal}{JETP letters} \textbf{\bibinfo{volume}{94}},
  \bibinfo{pages}{693} (\bibinfo{year}{2012}).

\bibitem[{\citenamefont{Cole et~al.}(2016)\citenamefont{Cole, Sau, and
  Das~Sarma}}]{cole2016proximity}
\bibinfo{author}{\bibfnamefont{W.~S.} \bibnamefont{Cole}},
  \bibinfo{author}{\bibfnamefont{J.~D.} \bibnamefont{Sau}}, \bibnamefont{and}
  \bibinfo{author}{\bibfnamefont{S.}~\bibnamefont{Das~Sarma}},
  \bibinfo{journal}{arXiv preprint arXiv:1603.03780}  (\bibinfo{year}{2016}).

\bibitem[{\citenamefont{Pedrocchi and
  DiVincenzo}(2015)}]{pedrocchi2015majorana}
\bibinfo{author}{\bibfnamefont{F.~L.} \bibnamefont{Pedrocchi}}
  \bibnamefont{and} \bibinfo{author}{\bibfnamefont{D.~P.}
  \bibnamefont{DiVincenzo}}, \bibinfo{journal}{arXiv preprint arXiv:1505.03712}
   (\bibinfo{year}{2015}).

\bibitem[{\citenamefont{Karzig et~al.}(2013)\citenamefont{Karzig, Refael, and
  von Oppen}}]{karzig2013boosting}
\bibinfo{author}{\bibfnamefont{T.}~\bibnamefont{Karzig}},
  \bibinfo{author}{\bibfnamefont{G.}~\bibnamefont{Refael}}, \bibnamefont{and}
  \bibinfo{author}{\bibfnamefont{F.}~\bibnamefont{von Oppen}},
  \bibinfo{journal}{Phys. Rev. X} \textbf{\bibinfo{volume}{3}},
  \bibinfo{pages}{041017} (\bibinfo{year}{2013}).

\bibitem[{\citenamefont{Cheng et~al.}(2011)\citenamefont{Cheng, Galitski, and
  Das~Sarma}}]{cheng2011nonadiabatic}
\bibinfo{author}{\bibfnamefont{M.}~\bibnamefont{Cheng}},
  \bibinfo{author}{\bibfnamefont{V.}~\bibnamefont{Galitski}}, \bibnamefont{and}
  \bibinfo{author}{\bibfnamefont{S.}~\bibnamefont{Das~Sarma}},
  \bibinfo{journal}{Phys. Rev. B} \textbf{\bibinfo{volume}{84}},
  \bibinfo{pages}{104529} (\bibinfo{year}{2011}).

\bibitem[{\citenamefont{Scheurer and
  Shnirman}(2013)}]{scheurer2013nonadiabatic}
\bibinfo{author}{\bibfnamefont{M.~S.} \bibnamefont{Scheurer}} \bibnamefont{and}
  \bibinfo{author}{\bibfnamefont{A.}~\bibnamefont{Shnirman}},
  \bibinfo{journal}{Phys. Rev. B} \textbf{\bibinfo{volume}{88}},
  \bibinfo{pages}{064515} (\bibinfo{year}{2013}).

\bibitem[{\citenamefont{Karzig et~al.}(2015{\natexlab{a}})\citenamefont{Karzig,
  Rahmani, von Oppen, and Refael}}]{karzig2015optimal}
\bibinfo{author}{\bibfnamefont{T.}~\bibnamefont{Karzig}},
  \bibinfo{author}{\bibfnamefont{A.}~\bibnamefont{Rahmani}},
  \bibinfo{author}{\bibfnamefont{F.}~\bibnamefont{von Oppen}},
  \bibnamefont{and} \bibinfo{author}{\bibfnamefont{G.}~\bibnamefont{Refael}},
  \bibinfo{journal}{Phys. Rev. B} \textbf{\bibinfo{volume}{91}},
  \bibinfo{pages}{201404} (\bibinfo{year}{2015}{\natexlab{a}}).

\bibitem[{\citenamefont{Karzig et~al.}(2015{\natexlab{b}})\citenamefont{Karzig,
  Pientka, Refael, and von Oppen}}]{karzig2015shortcuts}
\bibinfo{author}{\bibfnamefont{T.}~\bibnamefont{Karzig}},
  \bibinfo{author}{\bibfnamefont{F.}~\bibnamefont{Pientka}},
  \bibinfo{author}{\bibfnamefont{G.}~\bibnamefont{Refael}}, \bibnamefont{and}
  \bibinfo{author}{\bibfnamefont{F.}~\bibnamefont{von Oppen}},
  \bibinfo{journal}{Phys. Rev. B} \textbf{\bibinfo{volume}{91}},
  \bibinfo{pages}{201102} (\bibinfo{year}{2015}{\natexlab{b}}).

\bibitem[{\citenamefont{Sau et~al.}(2012)\citenamefont{Sau, Tewari, and
  Das~Sarma}}]{sau2012experimental}
\bibinfo{author}{\bibfnamefont{J.~D.} \bibnamefont{Sau}},
  \bibinfo{author}{\bibfnamefont{S.}~\bibnamefont{Tewari}}, \bibnamefont{and}
  \bibinfo{author}{\bibfnamefont{S.}~\bibnamefont{Das~Sarma}},
  \bibinfo{journal}{Phys. Rev. B} \textbf{\bibinfo{volume}{85}},
  \bibinfo{pages}{064512} (\bibinfo{year}{2012}).

\bibitem[{\citenamefont{Groth et~al.}(2014)\citenamefont{Groth, Wimmer,
  Akhmerov, and Waintal}}]{groth2014kwant}
\bibinfo{author}{\bibfnamefont{C.~W.} \bibnamefont{Groth}},
  \bibinfo{author}{\bibfnamefont{M.}~\bibnamefont{Wimmer}},
  \bibinfo{author}{\bibfnamefont{A.~R.} \bibnamefont{Akhmerov}},
  \bibnamefont{and} \bibinfo{author}{\bibfnamefont{X.}~\bibnamefont{Waintal}},
  \bibinfo{journal}{New Journal of Physics} \textbf{\bibinfo{volume}{16}},
  \bibinfo{pages}{063065} (\bibinfo{year}{2014}).

\bibitem[{\citenamefont{Sachdev}(2011)}]{sachdev2011quantum}
\bibinfo{author}{\bibfnamefont{S.}~\bibnamefont{Sachdev}},
  \emph{\bibinfo{title}{Quantum phase transitions}}
  (\bibinfo{publisher}{Cambridge University Press}, \bibinfo{year}{2011}).

\bibitem[{\citenamefont{{Olexei Motrunich, Kedar Damle, and David A.
  Huse}}(2001)}]{Dam}
\bibinfo{author}{\bibnamefont{{Olexei Motrunich, Kedar Damle, and David A.
  Huse}}}, \bibinfo{journal}{Phys. Rev. B.} \textbf{\bibinfo{volume}{63}},
  \bibinfo{pages}{224204} (\bibinfo{year}{2001}).

\bibitem[{\citenamefont{Datta}(1997)}]{datta1997electronic}
\bibinfo{author}{\bibfnamefont{S.}~\bibnamefont{Datta}},
  \emph{\bibinfo{title}{Electronic transport in mesoscopic systems}}
  (\bibinfo{publisher}{Cambridge university press}, \bibinfo{year}{1997}).

\bibitem[{\citenamefont{Willet}(2013)}]{willet2013qh}
\bibinfo{author}{\bibfnamefont{R.}~\bibnamefont{Willet}},
  \bibinfo{journal}{Reports on Progress in Physics}
  \textbf{\bibinfo{volume}{76}}, \bibinfo{pages}{076501}
  (\bibinfo{year}{2013}).

\bibitem[{\citenamefont{Das~Sarma
  et~al.}(2015{\natexlab{b}})\citenamefont{Das~Sarma, Hui, Brydon, and
  Sau}}]{sarma2015substrate}
\bibinfo{author}{\bibfnamefont{S.}~\bibnamefont{Das~Sarma}},
  \bibinfo{author}{\bibfnamefont{H.-Y.} \bibnamefont{Hui}},
  \bibinfo{author}{\bibfnamefont{P.}~\bibnamefont{Brydon}}, \bibnamefont{and}
  \bibinfo{author}{\bibfnamefont{J.~D.} \bibnamefont{Sau}},
  \bibinfo{journal}{New Journal of Physics} \textbf{\bibinfo{volume}{17}},
  \bibinfo{pages}{075001} (\bibinfo{year}{2015}{\natexlab{b}}).

\bibitem[{\citenamefont{Lutchyn et~al.}(2011)\citenamefont{Lutchyn, Stanescu,
  and Das~Sarma}}]{lutchyn2011search}
\bibinfo{author}{\bibfnamefont{R.~M.} \bibnamefont{Lutchyn}},
  \bibinfo{author}{\bibfnamefont{T.~D.} \bibnamefont{Stanescu}},
  \bibnamefont{and}
  \bibinfo{author}{\bibfnamefont{S.}~\bibnamefont{Das~Sarma}},
  \bibinfo{journal}{Physical review letters} \textbf{\bibinfo{volume}{106}},
  \bibinfo{pages}{127001} (\bibinfo{year}{2011}).

\bibitem[{\citenamefont{Stanescu et~al.}(2011)\citenamefont{Stanescu, Lutchyn,
  and Das~Sarma}}]{stanescu2011majorana}
\bibinfo{author}{\bibfnamefont{T.~D.} \bibnamefont{Stanescu}},
  \bibinfo{author}{\bibfnamefont{R.~M.} \bibnamefont{Lutchyn}},
  \bibnamefont{and}
  \bibinfo{author}{\bibfnamefont{S.}~\bibnamefont{Das~Sarma}},
  \bibinfo{journal}{Physical Review B} \textbf{\bibinfo{volume}{84}},
  \bibinfo{pages}{144522} (\bibinfo{year}{2011}).

\bibitem[{\citenamefont{Blonder et~al.}(1982)\citenamefont{Blonder, Tinkham,
  and Klapwijk}}]{blonder1982transition}
\bibinfo{author}{\bibfnamefont{G.~E.} \bibnamefont{Blonder}},
  \bibinfo{author}{\bibfnamefont{M.}~\bibnamefont{Tinkham}}, \bibnamefont{and}
  \bibinfo{author}{\bibfnamefont{T.~M.} \bibnamefont{Klapwijk}},
  \bibinfo{journal}{Phys. Rev. B} \textbf{\bibinfo{volume}{25}},
  \bibinfo{pages}{4515} (\bibinfo{year}{1982}).

\end{thebibliography}

\end{document}